\documentclass[english,fleqn,allpages]{llncs}

\usepackage{iste19} 

\usepackage{amssymb} 
\usepackage[cmex10]{amsmath}
\usepackage{array}
\usepackage{url}
\usepackage{cite}
\usepackage{color}

\usepackage{proof}
\usepackage{graphicx}
\usepackage{listings}
\usepackage{coq}
\usepackage{todonotes}

\begin{document}

\title{Computational Logic for Biomedicine and Neuroscience}

\author{
   Elisabetta De Maria\inst{1} \and Jo{\"e}lle Despeyroux\inst{2} \and Amy Felty \inst{3}
  \and Pietro Li{\`o} \inst{4} \and Carlos Olarte \inst{5} \and Abdorrahim Bahrami\inst{3} }

\institute{
I3S Laboratory, Sophia-Antipolis, France,  
\and
INRIA and CNRS, I3S Laboratory, Sophia-Antipolis, France
\and
School of Electrical Engineering and Comp. Science, University of Ottawa,  Canada
\and
Department of Computer Science and technology, University of Cambridge, UK 
\and
ECT -- Universidade Federal do Rio Grande do Norte, Brazil
}

\maketitle

\begin{abstract} 
We advocate here the use of computational logic for systems biology, as a \emph{unified and safe} 
framework well suited for both modeling the dynamic behaviour of biological systems,
expressing properties of them, and verifying these properties.
The potential candidate logics should have a traditional proof theoretic pedigree
(including either induction, or a sequent calculus presentation enjoying cut-elimination and focusing),
and should come with certified proof tools.
Beyond providing a reliable framework, this allows the correct encodings of our biological systems. 
For systems biology in general and biomedicine in particular, 
we have so far, for the modeling part, three candidate logics: all based on linear logic.
The studied properties and their proofs are formalized in a very expressive (non linear) inductive logic:
the Calculus of Inductive Constructions (CIC).
The examples we have considered so far are relatively simple ones;
however, all coming with formal semi-automatic proofs in the Coq system, which implements CIC.
In neuroscience, we are directly using CIC and Coq, 
to model neurons and some simple neuronal circuits and prove some of their dynamic properties.
In biomedicine, the study of multi omic pathway interactions, together with clinical and
electronic health record data should help in drug discovery and disease diagnosis.
Future work includes using more automatic provers.
This should enable us to specify and study more realistic examples,
and in the long term to provide a system for disease diagnosis and therapy prognosis.
\end{abstract}

\section{Introduction} \label{sec:lf-intro}
To formally model biological systems, several approaches have been proposed in the literature.
Discrete and hybrid Petri nets \cite{ChaouiyaNalRremyThie:11NC, HofestThe:98isb}
 are directed-bipartite graphs composed of places and transitions. In
$\pi$-calculus \cite{RegevSilSha:01psb} and its stochastic variant \cite{PhilipsCardelli:05tcsb},
processes communicate on complementary channels identified by specific names.
Bio-ambients \cite{RegevPpanSilCcarSha:04tcs}
are based on bounded places where processes are contained and where communications take place.
Hybrid automata \cite{ABIKMPRS01hscc}
combine finite state automata with continuously evolving variables.
Piecewise linear equations \cite{JongGouzeHPSG-04} capture the switch-like interactions between components.
Rule-based modeling languages such as Biocham \cite{FagesSolCha:04jbpc, Fages-Martinez-Rosenblueth-Soliman-2018},
Kappa \cite{DanosLaneve:04-tcs, BoutillierMLMKF-2018-bioinformatics},
and BioNetGen \cite{BFGH04bi} describe how (sets of) reactants can be transformed into (sets of) products, and associate corresponding rate-laws. They offer discrete, stochastic, or continuous semantics.
To express the dynamics of the different components, ordinary and stochastic differential equations are massively used too.
Answer Set Programming (ASP) is an unusual constraint logic programming approach 
enabling the modeling and study of large-scale biological systems.
Molecular Logic \cite{Wynn-Consul-Merajver-Schnell-2012} uses boolean logic gates
to define regulations in networks. 
Finally,
one of the most successful approaches to model and analyse signal transduction networks, inside the cells,
is Pathway Logic \cite{Talcott-Dill:tcsb06}: a system based on rewriting rules.

One of the most common approaches to the formal verification of biological systems is
symbolic model checking \cite{Clarke:2018book},
which exhaustively, but possibly by implicit boolean formulas,
enumerates all the states reachable by the system.
In order to apply such a technique, the biological system should be encoded as a finite transition
system and relevant system properties should be specified using  temporal logic.
Other approaches simply use simulators.
In addition to its simulator, Kappa offers powerful static -and causal- analysis of the models
\cite{DanosFFK-08-vmcai, BoutillierCCFLT-2018-cmsb, DanosFFHK:07-concur, DanosFFHHKTW-12-fsttcs}. 

In contrast to the aforementioned approaches, our approach is an unified approach, 
in which logic is used both to model biological systems, express their temporal properties and prove these properties.
Depending on the case, the logics chosen may be different. However, they are always highly expressive, 
\emph{computational} logics that can be used as \emph{logical frameworks}.

A \emph{computational logic} is a logic that enables proofs on the computer,
i.e. mechanized reasoning: one of the foundations of Artificial Intelligence (AI).
Formal logic emerged from philosophy in the 19th century, 
in an attempt to understand and formalize mathematical reasoning. 
The field was then further developed by mathematicians and more recently by computer scientists.
Various logics have been proposed.
From propositional logic to linear or inductive logics, 
the more expressive the logic, the less automated proofs can be.
Modern proof assistants, however, enable partially automated proofs.

\emph{Logical frameworks} are logics designed to formally study a variety of systems, 
such as transition systems, semantics of programming languages, mathematical theories, or even logics. 
Logical frameworks allow both the formal modeling of these systems 
and the proof of properties of the systems at hand. 
In the case where the logical systems are themselves logics, the logical framework
enables the proofs of both
\emph{meta-theoretical} theorems (about the logic being formalised) and
\emph{object level} theorems (about the systems being encoded in the formalised logic).
We shall follow this approach (sometimes called ``the two-level approach'') in the first part of this chapter (Sec. \ref{sec:lf-biomed}).

We advocate the use of computational logic 
as an \emph{unified and safe} approach to both specifying and analysing biological systems.
Furthermore, as we shall see in Sec. \ref{sec:lf-biomed}, our approach 
is general and can be used to model biological networks both inside and outside the cell. 

This chapter presents two applications areas of our approach: 
biomedicine (Sec. \ref{sec:lf-biomed}) and neuroscience (Sec. \ref{sec:lf-neuro}).
The logics used for modeling the systems will be different in these two areas.
On the other hand, the properties will be written and the proofs realized in the same logic.

In the first part of this chapter (Sec. \ref{sec:lf-biomed}), devoted to biomedicine, 
biological systems will be modeled in linear logic (see Sec. \ref{sec:ll}): 
a logic well suited for the studying of discrete state transition systems.

In the second part of the chapter (Sec. \ref{sec:lf-neuro}), devoted to an application to neuroscience,
neurons (and sets of neurons) will be modeled directly in the 
Calculus of Inductive Constructions (CIC): a general, typed and inductive logic.
More precisely, neurons will be described in the Coq system, used here 
as a high-level typed functional language 
which proves to be well suited to the description of neurons, their potential function and their combinations. 
Coq includes a programming language (Gallina). 
Coq is also, above all, a proof assistant \cite{BertotCasteran:2004}, which implements CIC.

In both cases, the properties will be written and the proofs realized in the same logic: CIC.
However, the properties will be written using CIC in a different way.
In the first case (biomedicine, Sec. \ref{sec:lf-biomed}), CIC and LL will be used in a two-level approach 
- as suggested above and described later (see Sec. \ref{sec:ccind-ll}).
In the second case (neurosciences, Sec. \ref{sec:lf-neuro}), CIC will be used directly. 
We shall use Coq to prove the properties of our systems 
in the two applications areas presented in this chapter.

The approach presented in this chapter is new, only proposed in four recent works: 
on the one hand \cite{DeMaria-Despeyroux-Felty:14-fmmb,DBLP:journals/tcs/OlarteCFH16,Despeyroux-Felty-Lio-Olarte:mlcsb18}
in biomedicine and on the other hand \cite{BDF18CSBIO} in neuroscience.
Here we choose discrete modeling. 
We believe that discrete modeling is crucial in systems 
biology since it allows taking into account some events 
that have a very low chance of happening (and could thus be neglected by
differential approaches), but which may have a strong impact on system behaviour.

\begin{figure} 
\begin{center}
\vskip -2.0cm
\includegraphics[scale=.70]{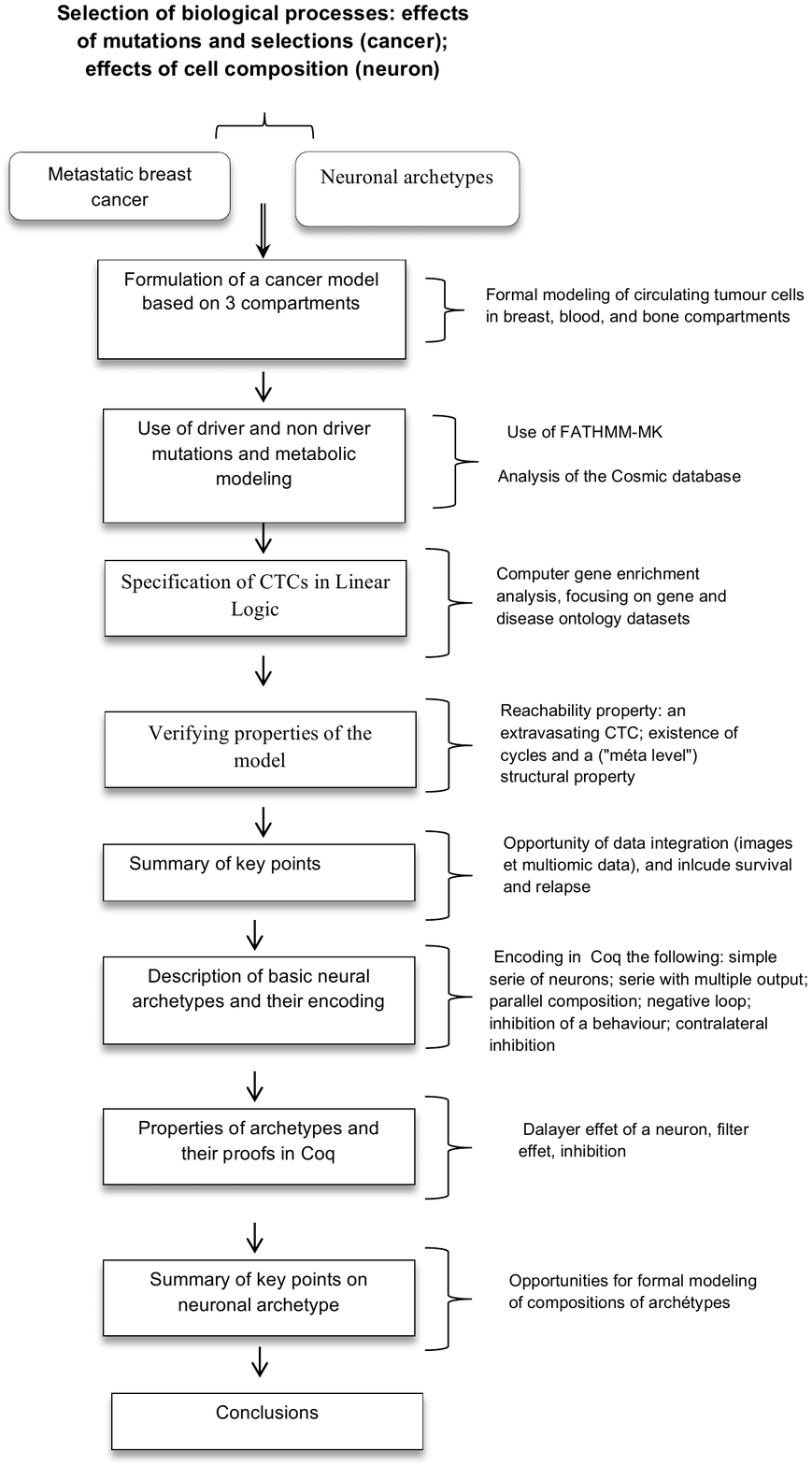}
\end{center}
\vskip -2.0cm \caption{Contents of the chapter.}
\label{fig:contents}
\end{figure}

The present chapter is made of two different models: metastatic breast cancer and neural archetypes.
These together make a good exemplification of modeling different aspects of biological complexity: 
cancer is about tissue cells becoming independent and neuron is a cell that builds circuits with other neurons. 
Fig. \ref{fig:contents} might help the reader understanding the structure of the chapter.

\section{Biomedicine in Linear Logic} \label{sec:lf-biomed} 

In this section, we present the use of logical frameworks, and more precisely linear logic, 
to model and analyse biological systems in the biomedicine area.
We focus here on metastatic breast cancer \cite{Despeyroux-Felty-Lio-Olarte:mlcsb18},
which is representative of our approach. 

We shall see that,
in contrast to usual mathematical approaches, which use differencial equations
to define, typically, an average number of cells,
in linear logic, it is more natural (but not required), to define single cell models.


\subsection{Introduction} \label{sec:biomed-intro}

Cancer is a complex evolutionary phenomenon, characterised by multiple
levels  of  heterogeneity (interpatient,  intrapatient  and intratumour), multiscale events (i.e. changes at intracellular,
intercellular, tissue levels), multiomics variability (i.e. changes to chromatine, epigenetic and transcriptomic levels) that affect all aspects of clinical decisions and practice. The most remarkable phenomenon is the occurrence of numerous somatic mutations, of which only a subset contributes to
cancer progression.
The dynamic genetic diversity, coupled with epigenetic plasticity, within each individual cancer induces new genetic architectures and clonal evolutionary trajectories. 

A striking feature of cancer is the subclonal genetic diversity, i.e., the presence of clonal succession and spatial segregation of subclones in primary sites and metastases. 
At the root of the subclonal expansion there are developmentally regulated potentially self-renewing cells. 
After initiation, multiple subclones often coexist, signalling parallel evolution with no selective sweep or clear fitness advantage that becomes evident with therapies. Fitness calculation for each subclonal is  difficult as subclones can form ecosystems and can cooperate through paracrine loops or interact through stromal, endothelial  and inflammatory cells. 

Tumour stages describe the progress of the tumour cells. One widely adopted approach is the American Joint Committee on Cancer (AJCC) tumour node metastasis (TNM) staging system. It classifies tumours with a combined stage between I-IV using three values: T gives the size of the primary tumour and extent of invasion, N describes if the tumour has spread to regional lymph nodes and M is indicative of 
distant metastasis. 
In terms of prognosis, Stage I patients have the best prognosis with 5 year survival rates (80-95\%). The survival rates progressively worsen with each stage. Even with advances in targeted therapies, Stage IV patients have survival rates of just over two years. 
The integration of blood tests, biopsies, medical imaging, with genomic data have allowed the classification of many subtypes of cancers with striking differences of driver mutations and survival patterns.
Therefore, the current data stream goals for a personalised breast cancer program should include the generation of tumour whole genome sequencing (DNA and RNA).
Another data stream goal could focus on liquid biopsies. This will consist of data obtained from circulating tumour DNA (ctDNA) and single cell analysis.
However, the full power of these datasets will not be
realised until we leverage advanced statistical, mathematical and computational approaches to devise the  needed procedures to conduct analyses that transect these streams. 
This condition clearly depends on the availability of a large amount of good quality data.

There is a very rich body of biomedical statistics, machine learning  and epidemiological literature for cancer data analysis which includes methods ranging from survival analysis, i.e., the effect of a risk factor or treatment with respect to cancer progression, analyses of co-alteration and mutual exclusivity patterns for genetic alterations, gene expression analyses, to network science algorithms (see e.g., \cite{AscolaniOcchipintiLio2015,Caravagna2018,Savage2012a,knutsdottir,enderling,gavaghan2002breast,bellomo2000modeling}). For example, in survival modeling, the data is referred to as the time to event date  and the objective is to analyse the time that passes before an event occurs due to one or more covariates \cite{Iuliano2016,Iuliano2018}. 
We believe that together with machine learning and biostatistics, 
there is a role for a logical approach in guiding
optimal treatment decisions and in developing a risk stratification and monitoring tool to manage cancer. 

Algorithms and benchmarks developed in cancer medicine
need to go through stages of standardisation and validations to become actionable
software and be used in worldwide clinical settings. This is a long multi-phase trajectory. 
Logic has the required interpretability, explanability, compositional and effectiveness 
to integrate data and protocols at different biomedical scales (single cell to human life style) 
in a monitorable and rigorous way. 

In this work, we focus on the use of a formal logical framework to provide  a reliable hypothesis-driven 
decision making system based on molecular data (single cell level). 
This first step is important in the investigation of the cancer to justify treatment.

The rest of this section is organised as follows. 
Section \ref{sec:ll} introduces Linear Logic. 
Section \ref{sec:ccind-ll} presents the ``two-level'' modeling approach we advocate. 
Section \ref{sec:cancer} first describes some relevant properties related to cancer mutations 
which we believe are key factors driving the model dynamics. 
Then we present our model of breast cancer progression. 
The formal proofs of some properties of our model are presented in Sec. \ref{sec:cancer-proofs}.
Finally we conclude with a discussion on challenges and opportunities of logical frameworks in cancer studies, and in biomedicine in general.
The reader may find the proofs of the results presented here in \cite{Despeyroux-Felty-Lio-Olarte:mlcsb18}.
Moreover, all the proofs of the properties of our model were certified in Coq
and are  available at \url{http://subsell.logic.at/bio-CTC/}. 

\subsection{Logical Frameworks, Linear Logic} \label{sec:ll} 

Among the many frameworks that have  been 
proposed in the literature,  linear logic~\cite{girard87tcs} (LL) is one of the most successful ones. This is mainly because   LL is resource conscious and, at the same time,  it can internalise \emph{classical} and \emph{intuitionistic} behaviours
(see, for example,~\cite{MillerPimentel:tcs13,cervesato02ic}).
\emph{Classical} logics, in which the law of excluded middle 
(which states that for any proposition, either that proposition is true or its negation is true) 
is valid, are natural to mathematicians,
while computer scientists - or mathematicians - doing mechanized proofs (on the computer) generally prefer intuitionistic logics.
In \emph{intuitionistic} logic, the law of excluded middle is not valid.
Moreover, any proof of the existence of a term $x$ satisfying some property $P(x)$ must produce a witness for it (i.e. a $t$ term such as $P(t)$ is true). 
Hence, LL allows for a  declarative and straightforward specification of 
transition systems as we shall see in brief. 
In fact, LL has been successfully used to model such diverse systems as
Petri nets, process calculi, security protocols,
multiset rewriting, graph traversal algorithms, and games.

LL is  general  enough for specifying and verifying properties of a large number of systems.
However, in some cases,  the  object-level system  (the one  we are modeling) exhibits
some characteristics that may pose difficulties in the modeling task.
For instance, in a transition system, we may be interested in 
specifying constraints on the timing for a transition to happen 
or, it could be also the case that transitions are constrained according to the spatial location of the objects.
Although those characteristics can be indeed modelled in LL
(see \cite{Despeyroux-Olarte-Pimentel:entcs-17,Chaudhuri-Despeyroux-Olarte-Pimentel:mscs-19}),
we may ask whether there are suitable extensions of the framework
offering a more natural representation for those constraints/modalities. 

Extensions of LL,  or its intuitionistic version ILL~\cite{girard87tcs}, have been proposed in order to fill this gap, hence providing 
stronger logical frameworks that preserve the elegant properties of  linear logic as the underlying logic.
Two such extensions are
HyLL (Hybrid Linear Logic) \cite{ChaudhuriDespeyroux:14}, an extension of ILL,
and SELL (Subexponential Linear Logic)
\cite{DanosJoinetSchellinx:93,nigam09ppdp},
an extension of ILL and LL. 
These logics  have been  extensively used for specifying systems that
exhibit modalities such as temporal or spatial ones. 
The difference between HyLL and SELL relies on the way those modalities are handled.

In the following, we shall introduce LL and its extensions HyLL and \sell\ in an intuitive way
and with the level of detail needed to understand the applications in Sec. \ref{sec:cancer-specif}.
 The reader may find in \cite{girard87tcs,troelstra92csli,ChaudhuriDespeyroux:14,nigam09ppdp} 
 a more detailed account on the proof theory of those systems. For didactic purposes, we shall consider a set of states $S$ and transitions of the form $s \redi{} s'$where $s,s' \in S$.
 Incrementally, we will add more structure to the states, e.g., we can consider that $s$ and $s'$ above are multisets.
Moreover, we will impose (temporal and spatial) constraints on the transitions 
in order to introduce the extensions of LL. 

\subsubsection{Linear logic} 

Modern logics, especially computational logics, are defined 
by the set of their \emph{formulas}, the form of their \emph{sequents}
(\emph{``judgements''} stating that a formula is true under a set of hypotheses),
and the \emph{inference rules} that explain how to build proofs in the logic at hand.

In Intuitionistic Linear Logic (ILL), formulas are defined by the following grammar
(LL's formulas include those of ILL, as well as other formulas that we will not use in this chapter):
\begin{itemize} 
  \item Terms:
    $\begin{array}{ll} t, ... ::= & c ~|~ x ~|~ f(\vec{t}) \end{array}$
  \item Propositions: \\ 
    $\begin{array}{ll}
    \qquad\qquad A, B, ... ::= & p(\vec{t}) ~|~ A \limp B ~|~
                      A \otimes B ~|~ \mathbf{1} ~|~ 
                      A \mathbin{\&} B ~|~ \top ~|~ A \oplus B ~|~  \mathbf{0} \\ 
    \qquad\qquad          & ! A ~|~ \forall x. A ~|~ \exists x. A 
    \end{array}$
\end{itemize}

A \emph{term} $t$ is a constant $c$, a variable $x$, or a function of one or more terms $f(\vec{t})$.
The simplest form of a formula/\emph{proposition} is $p(\vec{t})$.
For example,
proteins $\cn{P53}$, phosphorylated $\cn{MAPK}$, and the complex $(\cn{TGF\beta},\cn{LTBP1})$
may be represented, respectively, by the terms 
$\cn{P53}$, $\cn{ph}(\cn{MAPK})$ and $\cn{complex}(\cn{TGF\beta},\cn{LTBP1}).$ 
Concentration, or the presence or absence of an element, will be represented, not by terms, but by propositions
-which may be true or false.
For  example: 
${\cal C}(\cn{P53,0.2})$, $\cn{pres}(x)$,  $\cn{abs}(y)$. 

One of the main features of LL is the distinction between 
a proposition always true -\emph{stable truth}- (example: ``Socrates is a man'') 
and a proposition considered as a \emph{resource}  (``I have one dollar'').
Any proposition $A$ is a resource;
$!A$ represents a resource that can be used/consumed an arbitrary number of times
- i.e. a stable truth.
Therefore, "$A \limp B$" means "give me one $A$ and I will give you one $B$'',
while "$(!A) \limp B$" 
represents the usual implication "B is true whenever A is".
This awareness of resources leads to the existence of 
two versions of some constructors. 
Thus LL has two forms of "and", denoted $\otimes$ and $\mathbin{\&}$.
In $A \otimes B$ and $A \mathbin{\&} B$,
both resources/actions ($A$ and $B$) are available/possible.
However, in $A \otimes B$, both actions will be carried out,
while in $A \mathbin{\&} B$, only one of the two actions will be performed.
%
In our applications, a set of two elements/molecules $A$ and $B$ in a given state
will be represented by $A \otimes B$,
while a system containing two biological rules $r_1$ and $r_2$ will be represented,
in an \emph{asynchronous} semantics (where only one rule can be fired at a time), by $r_1 \mathbin{\&} r_2$.
Each of these conjunctions has its neutral element: $\mathbf{1}$ for $\otimes$ and $\top$ for $\mathbin{\&}$.
The connective $\oplus$ represents an "or''. Its neutral element is $\mathbf{0}$.
ILL has only one version of the ``or''; while LL (a perfectly symmetric logic) has two.
%
%
Finally, 
LL is a \emph{first-order logic}, whose formulas therefore include universal and existential quantifiers
on variables: $\forall x. A$ and $\exists x. A$.

The simplest form of a \emph{sequent} is $\Gamma \vdash G$
where $G$ is the formula (goal) to be proved (examples in Sec. \ref{sec:cancer-proofs}), 
with $\Gamma$ being a set of hypotheses (also formulas).
LL is a \emph{substructural} logic where there is an explicit control over
the number of times a formula can be used in a proof.
So-called ``Structural rules'', such as
contraction (which ignores the number of occurrences of hypotheses in $\Gamma$)
and weakening (which allows a hypothesis to be ignored),
usual in (non-linear) logic are not available here.
Formulas in LL can be  split into two sets: classical (those that can be used as many times as needed)
or linear (those that are consumed after being used). 
Using a dyadic system for LL, \emph{sequents} take the form  $\Gamma ~;~ \Delta \vdash G$
where $G$  is the formula (goal) to be proved, 
 $\Gamma$ is the set of classical formulas/hypotheses and $\Delta$ is the multiset of linear formulas/hypotheses. 
A proof of a $\Gamma ~;~ \Delta \vdash G$ judgment may or may not use the classical assumptions (in $\Gamma$). 
It must, on the other hand, use/consume all the resources in the linear context ($\Delta$).
We will use a \emph{focused} system \cite{DBLP:journals/logcom/Andreoli92} here.
By building what might be called a normal form of the proofs, 
such a system reduces the non-determinism during automatic proof search.
For that, we will decorate some of the formulas with arrows (see $\Downarrow A$ below) 
to unequivocally determine the next connective/formula that needs to be considered. 

In the linear context, classical formulas 
are marked with the exponential modality $!$,
whose left introduction rule (reading the inference rule from top to bottom, i.e. from the premises to the conclusion)
is as follows:
 \[
 \infer[!_L]{\Gamma ; \Delta , !F \vdash G}{\Gamma, F; \Delta  \vdash G}
 \]
This rule (reading from the conclusion to the premises, as usually done in proof-search procedures)
simply \emph{stores} the formula $F$ in the classical context $\Gamma$. 

In our applications, we shall store in  $\Gamma$  the formulas representing the rules of the system
 and in  $\Delta$ the atomic predicates (that can be produced and consumed) representing -a sub-part of- the state of the system.

For the moment, let us assume that $S$ is a finite set of states,
and each element $s\in S$ will be represented as an atomic proposition in LL also denoted as $s$.
A transition rule of the form $s \redi{} s'$  can be naturally specified as the linear implication $s \limp s'$
where, $s$ is consumed to later produce $s'$. The rules for this connective are:
\[
\infer[\limp_L]{\Gamma ; \Delta_1, \Delta_2,   \Downarrow F_1 \limp F_2 \vdash G}{
 \deduce{\Gamma ; \Delta_1   \vdash \Downarrow F_1}{}
 &
 \deduce{\Gamma ; \Delta_2, \Downarrow F_2   \vdash G}{}
}
\qquad
\infer[\limp_R]{\Gamma ; \Delta \vdash F_1 \limp F_2}{\Gamma ; \Delta , F_1 \vdash F_2}
\]

A left introduction rule, as $\limp_L$ above, explains how to \emph{use} a formula
(here $\Downarrow F_1 \limp F_2$) 
in a \emph{proof} (here simply a \emph{derivation tree}), that is built bottom up.
A right introduction rule, as $\limp_R$, describes how to \emph{prove} a formula.
Such a set of rules \emph{defines} the top operator (here $\limp$) of the given formula.

In $\limp_R$, the proof of $F_1 \limp F_2$ requires 
(in addition to the resources in $\Delta$ and possibly assumptions in $\Gamma$)
the use of the resource $F_1$ to conclude $F_2$. 
This rule is invertible (i.e., the premise is provable if and only if the conclusion is provable).
Hence, this rule belongs to the so-called \emph{negative} phase of the construction of a focused proof, 
where, without losing provability, we can apply all the invertible rules in any order.

The rule $\limp_L$ shows the resource awareness of the logic:
part of the context ($\Delta_1$) is used to prove $F_1$ 
and the remaining resources ($\Delta_2$) must be used (in addition to $F_2$) to prove $G$.
The classical context $\Gamma$ is not divided but copied in the premises.
The rule $\limp_L$ is  non-invertible  and then, it belongs to the so-called \emph{positive} phase.
This means that there may be several ways to prove the conclusion: for example 
by "decomposing" the top operator of the goal $G$, instead of decomposing the implication ($\limp$) in 
the resources/assumptions of the sequent-conclusion.
Therefore, 
if we decide to work/focus on that formula, it may be the case that we have to backtrack if the proof cannot be completed.  
The notation $\Downarrow F_1 \limp F_2$ (read $\Downarrow (F_1 \limp F_2)$)
means that we \emph{decided} to \emph{focus on} that formula
and then, we have to keep working on the sub-formulae $F_1$ and $F_2$ 
(notation $\Downarrow F_1$ and $\Downarrow F_2$). 
During the negative phase as during the positive phase, the context (linear or classical) can have zero, one or more focused formulas. In any case, one can only focus on one formula.

As an example, in $\Gamma$ above,  we can store formulas of the shape $s \limp s'$ specifying the transition rules of the system (that can be used as many times as needed).
Assuming that all the negative connectives have been already introduced, 
the following derivation shows how to focus on / select a formula $s \limp s'$ stored in $\Gamma$.
This is the purpose of the decision rule $D_C$: 
\[
\infer[D_C]{\Gamma, s\limp s' ; \Delta \vdash G}{
   \Gamma, s\limp s' ; \Delta , \Downarrow s\limp s' \vdash G
 }
\]

Note that the rule $D_C$ creates a copy of the formula $s \limp s'$ and places it in the linear context. 

If the current (partial) state $\Delta$ contains the formula $s$, the proof can proceed by applying rule $\limp_L$,
meaning applying the biological rule $s \limp s'$:
\[
\infer[D_C]{\Gamma, s\limp s' ; s, \Delta \vdash G}{
\infer[\limp_L]
  {\Gamma, s\limp s' ; s, \Delta, \Downarrow s \limp s' \vdash G}
  {
   \infer[I]{\Gamma, s\limp s' ; s  \vdash \Downarrow  s}{}
   &
   \deduce{\Gamma, s\limp s' ; \Delta, \Downarrow s'  \vdash G}{}
  }
 }
\]

$I$ is the initial rule that allows us to prove an atomic proposition 
if such atomic proposition is the unique formula in the linear context
(alternatively, the atomic proposition can be in the classical context $\Gamma$ and the linear context must be empty).
When looking at this derivation, bottom up, we observe that 
the multiset $\{s, \Delta \}$ was transformed into $\{\Delta, \Downarrow s' \}$, 
thus reflecting  (almost completely) the transition ($s \limp s'$) in the biological system (the "focus" on $s'$ remains to be removed).
We will see this in more detail, considering a more general form of biological rule.

Usually, in biochemical systems, the state of the system is composed of a (multi) set of components.
Rules describe how one or more reactants are consumed in order to produce some other components.
For instance, a typical rule may be 
"$\textrm{cdk}46 + \textrm{cycD} \redi{} \textrm{cdk46-cycD}$" 
representing cdk binding to a cyclin, in the cell cycle.
%

In Linear Logic, the transition above will be represented by the formula
``$\textrm{cdk}46 \otimes \textrm{cycD} \limp \textrm{cdk46-cycD}$".
This representation requires the use of the conjunction $\otimes$ 
and the following logical rules defining it: 
\[
 \infer[\otimes_L]{\Gamma ; \Delta, F_1 \otimes F_2 \vdash G}{\Gamma ; \Delta, F_1, F_2 \vdash G}
 \qquad
 \infer[\otimes_R]{\Gamma;\Delta_1, \Delta_2 \vdash \Downarrow F_1 \otimes F_2}{
 \deduce{\Gamma;\Delta_1 \vdash \Downarrow F_1}{}
 &
 \deduce{\Gamma;\Delta_2 \vdash \Downarrow F_2}{}
 }
\]

The rule $\otimes_R$ belongs to the positive phase and it says that the proof of $F_1 \otimes F_2$ requires the linear context to be split in order to prove both $F_1$ and $F_2$.
The left rule ($\otimes_L$) belongs to the negative phase and 
the resource $F_1 \otimes F_2$ is simply transformed into two resources ($F_1$ and $F_2$).
For instance, assuming that $A,B,C,D$ are propositional variables and letting $F = (A \otimes B) \limp  (C \otimes D)$ and $F\in \Gamma$, we obtain the following derivation: 
\[
\infer[\limp_R]{\Gamma ; \Delta \vdash (A \otimes B) \limp G}{
 \infer[\otimes_L]{\Gamma ; \Delta, A \otimes B \vdash  G}{ 
  \infer[D_C]{\Gamma ; \Delta, A, B \vdash  G}{
   \infer[\limp_L]{\Gamma ; \Delta , A, B, \Downarrow F \vdash G}{
     \infer[\otimes_R]{\Gamma ; A,B \vdash \Downarrow (A \otimes B)}{
      \infer[I]{\Gamma;A \vdash \Downarrow A}{}
      &
      \infer[I]{\Gamma;B \vdash \Downarrow B}{}
     }
     &
     \infer[R]{\Gamma;\Delta, \Downarrow(C\otimes D) \vdash G}{
      \infer[\otimes_L]{\Gamma ;\Delta, C \otimes D \vdash G}{\Gamma ;\Delta, C, D \vdash G}
     }
   }
  }
 }
}
\]

Here we have introduced a new rule. 
The rule $R$, for release, means that the focused phase is finished since the formula $C\otimes D$ (on the left) belongs to the negative phase.
When looking at this derivation, bottom up, we observe that in a switch of the polarity of the proof
(a complete positive phase followed by a negative one)
the multiset of components $\{\Delta, A, B\}$ was transformed into $\{\Delta, C,D\}$, 
thus reflecting the transition ($F$) in the biological system. 
This derivation is a partial proof, in which it remains to prove 
the goal $G$, with (the assumptions in $\Gamma$ if needed and) the new resources $\{\Delta, C, D\}$.

The attentive reader will have noticed that we have not given the definition of all the constructors of ILL. 
Some inference rules -not necessary for the understanding of the applications presented in this chapter- 
have been deliberately omitted in order to simplify the presentation of the logic. 

A few words on syntax versus semantics.
We said that the constructors of a logic were defined by their introduction/elimination rules.
This is correct, in a \emph{syntactic} approach/definition of logic. 
The syntactic approach is the usual approach in computational logic: computers only understand syntax.
Nevertheless, even in computational logic, it can be interesting to define the \emph{semantics} of a logic,
to better understand her.
This only makes sense for a very expressive logic. 
Otherwise, semantics are trivial, like, typically, truth tables for propositional logic: 
"$A \land B$ is true if $A$ is true and $B$ is true" seems like a truism ("Lapalissade")...
LL thus has a semantics that represents proofs by hypergraphs: proof nets.
Proof nets allow, notably, to identify two derivations  
syntactically different, although "morally"/semantically identical (e.g. except for rule permutations).

\subsubsection{Adding modalities} \label{sec:modalities-adequacy}
One may want to consider adding the time needed for a transition to happen.
Let us then consider a transition of the form $s \transR{d} s'$ which means that the state $s$ is transformed into $s'$ in $d$ time-units.
For the sake of simplicity, we consider here that $d$ is a constant; however, it can also be a function
(examples in Sec. \ref{sec:cancer-specif}).  
We may add to the model a unary predicate $t(\cdot)$ denoting the current time-unit and then, the transition above can be modelled by the following formula 
\begin{equation} \label{eq:forimp}
\forall n, (t(n)\otimes s)  \limp (t(n+d) \otimes s')
\end{equation}
The rules for the universal quantifier are the following 
\[
\infer[\forall_L]{\Gamma; \Downarrow \forall x. F \vdash G}{
\Gamma; \Downarrow F[t/x] \vdash G
}
\qquad
\infer[\forall_R]{\Gamma ;\Delta \vdash \forall x. G}{\Gamma ;\Delta \vdash  G[x_e/x]}
\]
where $x_e$ is a fresh variable (not occurring else where).
Note that $\forall_L$ belongs to the positive phase, where a term $t$ must be provided to continue the derivation.
On the other side, $\forall_R$ belongs to the negative phase and $x$ is simply replaced by a fresh variable.
Let $F$ be the formula \eqref{eq:forimp}, it is easy to complete the missing steps in the derivation below:

\[
\infer[D_C]{\Gamma ; \Delta, s, t(x) \vdash G}{
 \infer={\Gamma ; \Delta, s, t(x), \Downarrow F \vdash G}{
 {\Gamma ; \Delta, s', t(x+d)  \vdash G}{}
 }
}
\]

The double bars in the above derivation represents the whole positive phase 
(decomposing the connectives in the rules $\forall_L$, $\limp_L$ and $\otimes_R$) 
followed by the negative phase (loosing focusing on the rule $\otimes_L$).
Again, note that a focused step of the proof reflects exactly a transition in the system. 

\subsubsection{Adequacy} \label{sec:adequacy}
In the following results, we shall use $\os s \cs_x$ to denote the formula $s \otimes t(x)$ (i.e., the state $s$ at time-unit $x$).  
We shall also use $s \transR{(r,d)} s'$  to denote  that the system may evolve from state $s$ to state $s'$
by applying the rule $r$ that takes $d$ time-units. Hence, 
  $S_s=\{(s',r,d) \mid  s \transR{(r,d)} s' \}$
represents the set of possible transitions starting from $s$.
Moreover, with \texttt{system} we denote the encoding of the transition rules of the system. 
 We can show that all transitions in $S_s$ match exactly one focused derivation of the encoded system. More precisely, 

\begin{theorem}[Adequacy] \label{thm:adequacy}
Let $s$ be a state and $S_s=\{(s',r,d) \mid  s \transR{(r,d)} s' \}$. 
Then,   $(s',r,d)\in S_s$
iff focusing on the encoding of $r$ leads to the following derivation. 
\[
\infer={\systemF ~; \encCTLH{s}_t \vdashseq G}{\systemF ~; \encCTLH{s'}_{t+d} \vdashseq G}
\]
 Moreover, let $s,s'$ be two states. 
Then $s \transR{(r,d)} s'$ iff the sequent below is provable
\[\systemF  ; \cdot \vdashseq \encCTLH{s}_t   \limp \encCTLH{s'}_{t+d} \]
\end{theorem}

The above results allow us to use the whole positive-negative phase as macro rules in the logical system. 
Formally, we can show that the corresponding logical rule is admissible in the system, 
i.e., if the premise is provable then the conclusion is also provable. 
\begin{corollary}[Macro rules] \label{cor-mrules}
Assume that $s \transR{(r,d)} s'$. Then, the following macro rule is admissible:
\[
\infer[r]{\systemF ~; \Delta , \encCTLH{s}_t \vdashseq G}{\systemF ~; \Delta, \encCTLH{s'}_{t+d} \vdashseq G}
\]
\end{corollary}

\subsubsection{Modalities as worlds}

This section and the next one may be overlooked at a first reading.
We present there two extensions of Linear Logic that allow interesting modelings of modalities.
Indeed, these logics have been used to model and analyze small
biological systems in two initial experiments
\cite{DeMaria-Despeyroux-Felty:14-fmmb, DBLP:journals/tcs/OlarteCFH16}. 
However, these modal extensions are not necessary for understanding the rest of the chapter.

Hybrid Linear Logic  (HyLL)
is a conservative extension of ILL where the truth judgements
are labelled by worlds representing constraints on states and state transitions.
Judgements of HyLL are of the form
``$A$ is true at world $w$'', abbreviated as $A ~@~ w$.
Particular choices of worlds produce particular instances of HyLL,
e.g., $A~@~t$ can be interpreted as ``$A$ is true at time $t$''.
HyLL was first proposed in~\cite{ChaudhuriDespeyroux:14} and it has  been used
as a logical framework for specifying modalities as well as biological
systems~\cite{DeMaria-Despeyroux-Felty:14-fmmb}.

Worlds are given meaning via a 
\emph{constraint domain} $\cal W$, which is a monoid structure $\langle W, ., \iota\rangle$
and
the  \emph{reachability relation} on worlds $\preceq\ : W \times W$ is defined as  $u \preceq w$
iff there exists $v \in W$ such that $u . v = w$.
The identity world $\iota$ is the $\preceq$-initial and it is intended to represent the
lack of any constraints.
Thus, the ordinary first-order ILL  can be  embedded into any
instance of HyLL by setting all world labels to the identity.
A typical  example of a constraint domain is
$\mathcal{T} = \langle \N, +, 0\rangle$, representing instants of time.

Formulas in HyLL are constructed from atomic propositions, connectives of ILL and the following  two hybrid connectives:
\emph{satisfaction} ($\texttt{at}$), which states that a
proposition is true at a given world   and
\emph{localisation} ($\downarrow$), which binds a name for the current world where
the proposition is true.  The rules of these connectives are: 
\[
  \infer[{[\at~ L]}]{\Gamma ; \Delta, (A ~\at~ u) ~@~ v \vdashseq C ~@~ w} {\Gamma ; \Delta, A ~@~ u \vdashseq C ~@~ w}
\qquad  \infer[{[\at~ R]}] {\Gamma ; \Delta \vdashseq (A ~\at~ u) ~@~ w}{\Gamma ; \Delta \vdashseq A ~@~ u} 
\]
\[
  \infer[{[\downarrow L]}]{\Gamma ; \Delta, \downarrow u. A ~@~ v \vdashseq C ~@~ w}{\Gamma ; \Delta, A [v / u] ~@~ v \vdashseq C ~@~ w}
\qquad \quad
  \infer[{[\downarrow R]}]{\Gamma ; \Delta \vdashseq \downarrow u. A ~@~ w}{\Gamma ; \Delta \vdashseq A [w / u] ~@~ w}    
\]

The simple example above, representing cdk binding to a cyclin, in the cell cycle,
can be written in HyLL as follows, where we specify the delay $d$ for the reaction:
"$\textrm{cdk}46 + \textrm{cycD} ~\textrm{at}~ t
 \limp \textrm{cdk46-cycD} ~\textrm{at}~ (t+d)$." 

\subsubsection{Modalities as subexpoentials}

Linear logic with subexponentials (\sell) shares with LL all
its connectives except the exponentials:
instead of having a single pair of exponentials $\bang$ and $\quest$
(we will not use $\quest$ in this chapter), \sell\ may
contain as many \emph{subexponentials}~\cite{DanosJoinetSchellinx:93,nigam09ppdp},
written $\nbang{a}$  and $\nquest{a}$, as one needs.

A \sell\ system is specified by a \emph{subexponential signature} $\Sigma = \tup{I, \preceq,U}$, 
where $I$ is a set of labels, $U \subseteq I$  specifies which
subexponentials are unbounded
and $\preceq$ is a pre-order among the elements of $I$. 
Intuitively, $\nbang{a}F$ means that $F$ is marked with a given modality $a$.
The preorder $\preceq$ on the indices $I$ determines the probability relation.
For instance, $\nbang{a}F$ proves $\nbang{b}F$ whenever $b \preceq a$ 
(intuitively, $\nbang{a}$ is a stronger modality that can be used in a proof of the weaker modality $\nbang{b}$).
Moreover, $\nbang{b}F$ cannot prove $\nbang{a}$ if $b \prec a$ 
(weaker formulas  cannot be used during a prove of stronger ones). 
As shown in \cite{NigamOlartePimentel:concur-13,NigamOlartePimentel:tcs-17},
the formula 
$\nbang{a}{F}$  can be interpreted in several ways,
for instance,  it  may represent the fact that $F$ holds in the space location $a$ or that $F$ holds in the time-unit $a$.
Moreover, if $a$ is unbounded  (or classical), then $F$ can be used as many times as needed.
The rules for $\nbang{a}$ are the following:
 \[
 \infer[!_L]{\Gamma ; \Delta,  \nbang{a}F \vdash G}{ \Gamma, a: F ; \Delta \vdash G}
 \qquad \quad
 \infer[!_R]{\Gamma ; \cdot \vdash \Downarrow \nbang{a}{G}}{\Gamma^{\preceq a} ; \cdot \vdash {G}}
 \]

 In $!_L$, we simply ``store'' the formula $F$ in the context $a$.
 This rule belongs to the negative phase.
 Rule $!_R$ belongs to the positive case. Note that the linear context must be empty.
 Moreover, the resulting $\Gamma^{\preceq a}$ context only contains formulas of the form $\nbang{b}{F}$ such that $a \preceq b$.

Now consider a constrained transition of the form
\[
[A]_a + [B]_b \redi{} [C]_c
\]
meaning that the component $A$ must be located in the  \emph{space domain} $a$
(similarly for $B$) in order to produce  $C$ in the space domain $c$. 
 
The component $[A]_a$  can be adequately represented as the formula $\nbang{a}{A}$ and the reaction above becomes
$
(\nbang{a}{A} \otimes\nbang{b}{B}) \limp \nbang{c}{C}
$. 

It is also possible to enhance  the expressiveness of SELL with the subexponential quantifiers $\forallLoc$
(``for all locations'') and $\existsLoc$ (``there exists a location'')
\cite{NigamOlartePimentel:tcs-17}
given by the rules 
\[
 \infer[\forallLoc_L]{\Gamma ; \Delta, \Downarrow \forallLoc \typeloc{l_x}{a}. F \vdash G}
 {\Gamma ; \Delta, \Downarrow, G[l/l_x] \vdash G}
 \qquad
  \infer[\forallLoc_R]{\Gamma ;\Delta \vdash \forallLoc \typeloc{l_x}{a}. G}
 {\Gamma ; \Delta \vdash   G[l_{e}/l_x]}
\]
where   $l_e$ is fresh.
Intuitively, subexponential variables play a similar role as eigenvariables. 
The generic variable $\typeloc{l_x}{a}$ represents any subexponential,
constant or variable in the ideal of $a$. Hence $l_x$ can be substituted
by any  subexponential $l$ of type $b$ (i.e., $l:b$) if
$b\preceq a$.
We call the resulting system $\sellU$. 

By ordering the location in the subexpoential signature, we can create hierarchies of spaces.
Then, we may stipulate that a certain reaction occurs in all the spaces related to a given location $x$ as follows:
\[
\forallLoc \typeloc{l}{x}. \left( \nbang{l}{A} \otimes\nbang{l}{B} \limp \nbang{l}{C}\right)
\]

In this case, we observe the transition $A + B \redi{} C$ in all the space domains subordinate to $x$. 

As shown in \cite{DBLP:journals/tcs/OlarteCFH16},
it is also possible to provide a suitable subexponential signature to combine
both spatial and temporal modalities in the same framework. 

The reader may find in  
\cite{Despeyroux-Olarte-Pimentel:entcs-17,Chaudhuri-Despeyroux-Olarte-Pimentel:mscs-19}
a formal comparison of HyLL and SELL,  
and an adequate encoding of Temporal Logic in LL extended with fixpoints.
HyLL may be used to describe \emph{stochastic processes}
(based on variables with exponential distributions).
This approach has been used to encode the S$\pi$-calculus into HyLL \cite{ChaudhuriDespeyroux:14}.
However, only symbolic operations on rates were needed there.
Using this approach to model and analyse biological systems, for example, is still future work.
%
%
In our later experiments, such as the one described in Sec. \ref{sec:cancer},
we chose to use pure LL, and define a predicate to encode time. 
This approach might be less elegant;
however, it avoids the extra complication of copying the unused (not consumed) information to the
next time-unit (HyLL world or SELL subexponential),
thus benefiting from the usual compositional nature of the logic.

\subsection{Modeling in Linear Logic} \label{sec:ccind-ll}

As we have seen in the introduction of the present chapter (Sec. \ref{sec:lf-intro}),
we formalize both biological systems and their properties in logic,
and prove these properties in logic as well.
We shall use here Linear Logic (LL) to model biological systems (here the evolution of cancer cells)
and the Calculus of Inductive Constructions (CIC) to write their properties and prove them.
More precisely, in a ``two-level approach'', we shall use Linear Logic (LL) 
as the intermediate logic, formalised in CIC, which is a type theory  
implemented in the Coq Proof Assistant \cite{BertotCasteran:2004}.
We note that 
the Coq system has been (partially) proven correct in itself,
extensive meta-theoretical studies of LL are available
in the Coq system (see e.g., \cite{Xavier-Olarte-Reis-Nigam:entcs17}),
and our encoding of biological systems is \emph{adequate}
(Section \ref{sec:modalities-adequacy}).
This means that we prove that the formal model of the system 
correctly encodes (the transition system modeling) the intended biological system.
The approach is thus an unified approach, fully based on logic, and
a safe approach, as each step is proved correct, as far as it can be
(G{\"o}del's theorem).

We leverage our formalisation of LL in CIC 
to give  a natural and direct characterisation of the state transformations of Circulating Tumour Cells (CTCs). 
For instance, a rule describing the evolution of a cell $n$, in a region $r$, 
from a healthy cell (with no mutations) 
to a cell that has acquired a mutation $\tgfbeta$ 
can be modelled by the linear implication 
$ \cn{C}(n,r,[]) \limp \cn{C}(n,r,[\tgfbeta])$. 
This formula describes the fact that a state where a cell
$\cn{C}(n,r,[])$ is present 
can evolve into a state where $\cn{C}(n,r,[\tgfbeta])$ holds. 
If this transition takes a delay $d$, we will formalise it by adding a predicate $T$, as proposed in Sec. \ref{sec:modalities-adequacy}:
$ \cn{T}(t) \otimes \cn{C}(n,r,[]) \limp \cn{T}(t+d) \otimes \cn{C}(n,r,[\tgfbeta])$. 
More interestingly, the LL specification can be used  to prove some
desired properties of the system.
For instance, it is possible to prove
\emph{reachability properties}, i.e., whether the system can reach a given state 
or even more abstract (meta-level) properties such as checking \emph{all} the possible evolution paths
the system can take under certain conditions (Sect. \ref{sec:cancer-proofs}). 
Finally,  we attain a certain degree of automation in our proofs which
opens the possibility of  testing recent proposed hypotheses in the literature. \\

It is worth noting that  knowledge of logic, and in particular of LL, is undoubtedly useful, but 
not necessary for the biologist interested in formalizing and studying his system following the approach proposed here.
LL has been implemented in Coq, with some of its rule systems (focused or not), 
as well as tactics for developing proofs, by different experts 
(see, for example, \cite{Xavier-Olarte-Reis-Nigam:entcs17,FeltyMomigliano:JAR10,Mahmoud-Felty-2018-lsfa}),
once and for all. 
The biologist potentially interested in using our approach should be able to easily model his system in LL
by following one of the applications given as an example in this chapter 
or in the preliminary studies cited \cite{DeMaria-Despeyroux-Felty:14-fmmb,DBLP:journals/tcs/OlarteCFH16}.
Even better, we think in the near future, 
not, a priori, to develop our own interface to a "biology dedicated LL", but to propose
translations/compilations of models written in rule-based languages such as Biocham and Kappa
(presented in other chapters of this book) to corresponding models written in LL. 
The writing of properties of interest, in the form of sequents, could probably, in simple cases, be done by the biologist.
However, in the current state of science, this step, like the previous one (modeling), 
often requires a long dialogue between the biologist and the computer scientist, 
if only to estimate the possible contribution of the proposed approach to the problem 
and the most appropriate degree of abstraction. 
The final step -the proofs- will probably have to be performed by experts, here in mecanised proofs, on the computer, 
although the development of powerful tactics (by these same experts) should eventually make this step also affordable to non-experts.

\subsection{Modeling Breast Cancer Progression} \label{sec:cancer}

We first describes here (\ref{sec:ctc-mut}) some relevant properties related to cancer mutations
and CTCs which we believe are key factors driving the model dynamics. 
Then, in Section \ref{sec:cancer-specif}, we specify in LL the dynamics of CTCs.
Reachability, existence of cycles, and meta-level properties of the system 
will be proven in Section \ref{sec:cancer-proofs}.
 
\subsubsection{Tumour Cells in Metastatic Breast Cancer} \label{sec:ctc-mut}

In this section we first describe the mutations involved in cancer in general  
and then, we focus on the evolution of circulating tumour cells described in this work. 

\paragraph{\bf Cancer Mutations.} 
Cancer mutations can be divided into
drivers and non-drivers (or passengers).
Non-driver mutations may change the metabolic network and affect important cell-wide processes such as phosphorilation and methylation. They could initiate a cascade of changes that generate effective clonal heterogeneity.
The accumulation of evidence of clonal heterogeneity and the observation of the emergence
of drug resistance in clonal sub-populations suggest that mutations usually classified
as non-drivers may have an important role in the fitness of the cancer cell and in the
evolution and physiopathology of cancer.  Similarly, mutations that alter the metabolism
and the epigenetics may modify the fitness of the cancer cells.  
A meaningful way to identify drivers and passenger mutations is to use a statistical  estimator of the impact of mutations such as {FATHMM-MKL} and a very large mutation database such as Cosmic (\url{http://cancer.sanger.ac.uk/}) \cite{DBLP:journals/bioinformatics/ShihabRGMCDGC15}. 

The mutation process (causing tumour evolution) generates intra-tumour heterogeneity. The subsequent selection and Darwinian evolution (including immune escape) on intra-tumour heterogeneity is the key challenge in cancer medi\-cine. 
Those clones that have progressed more than the others will have larger influence on patient survival and determine the cancer subtype stratification of the patient. The amount of heterogeneity within primary tumours or between primary and metastatic can be huge. 
The heterogeneity could be investigated through molecular biology techniques such as single cell sequencing, in situ Polymerase Chain Reaction (PCR) 
and could also be phenotypically classified using microscope image analysis. 
In case of large heterogeneity we could assume that the survival of the patient strongly depends on the 
mutations of the most aggressive clones/cells.
Therefore single cell sequencing at primary site and metastatic sites could be highly informative of the level of heterogeneity; probably more than bulk tissue gene expression analysis. 
Although still used only in few clinical protocols, cancer single cell analysis may spread quickly  as it can be
easily integrated with other tissue, organ, blood and patient level (imaging, life style) information in a clinical decision system.
This observation motivates the choice of single cell models in this methodological study.

\paragraph{\bf Circulating Tumour Cells (CTCs).}
We follow the study of the evolution of Circulating Tumour Cells in
metastatic breast cancer  in \cite{AscolaniOcchipintiLio2015},
where the authors use differential equations. This reference has 
an extensive discussion
on the modeling choices, in particular concerning the driver mutations.

In \cite{AscolaniOcchipintiLio2015} the probability for a cell in a duct in the breast
to metastasise in the bone depends on the following mutational events:
\begin{enumerate}
\item{}
A mutation in the $\tgfbeta$ pathways frees the cell from the surrounding cells.

\item{} A mutation in the $\epcam$ gene makes the cell rounded and free to divide.
Then the cell enters the blood stream and becomes a circulating cancer cell.

\item{} In order to survive, this cell needs to over express
the gene $\cdfs$ that prevents attacks from the immune system. 

\item{} 
Finally,  there are two mutations that allow the circulating cancer cell to attach to 
the bone tissue and start the deadly cancer there: \cdff\ and \met.
\end{enumerate}

Hence, a cancer cell has four possible futures:
(a) acquiring a driver mutation;
(b) acquiring a passenger mutation which does not cause too much of a viability 
problem: it simply increases a sort of ``counter to apoptosis'';
(c), the new (i.e., last) mutation brings the cell to apoptosis;
and (d), moving to the next compartement, or seeding in the bone.

The behaviour of the cells depends on the \emph{compartments} the cells live in
 (here the breast, the blood and the bone),
the other cells (i.e., the \emph{environment}: 
the  availability of food/oxygen or the pressure by the other cells), 
and the behaviour of the surface proteins (the \emph{mutations}).
In this work, we shall formalise the compartments and the mutations, and 
leave the formalisation of the environment 
to future work. 
Note that this environment plays a role only in the breast.

The \emph{phenotype} of a cell is characterised by both the number of its mutations and its fitness. 
In biology the {\it fitness} is the capability of the cell to survive
and produce offsprings. The cell's viability is particularly dependent on metabolic health and energy level. 
Most of the cell's metabolic health depends on the accumulation of mutations that affect the production of enzymes involved in catalysing energy-intensive reactions and cell homeostasis. 
The fitness is particularly altered by the occurrence of driver mutations: each driver mutation provides the cell with additional fitness.
Non-driver mutations, on the other hand, may accumulate in large numbers, and may affect cell stress response due to the altered metabolism and the competition with other neighboring cells \cite{DiGregorio2016}.
Wet-lab tests for cell fitness and stress responsiveness have been recently developed, see for instance  \cite{Antczak2014,Venkataram2016,Rogers2017}.
In our formalisation, the fitness will be a parameter of the cells.
Physicians see the appearance of the cell (round, free, etc.), while biologists see the mutations;
our model can take both into account.

We extend the model in \cite{AscolaniOcchipintiLio2015} with a few rules
modeling DNA repair of passenger mutations.
These rules, only available for cells with $\tgfbeta$ or $\epcam$ mutations, (i.e. before $\cdfs$ mutation), 
represent DNA repair by increasing the fitness by one.
Note that this addition introduces \emph{cycles} in our model
(i.e., it is possible for a cell to go back to a previous state).

\subsubsection{Modeling CTCs in Linear Logic} \label{sec:cancer-specif}

In this section we formalise in LL the behaviours of the Circulating Tumour Cells. 
The correctness of this formalisation 
(with respect to the corresponding traditional modeling as a transition system)
was proven in a generic way in Sec. \ref{sec:ll}.

We have seen in Sec. \ref{sec:ll} that 
LL formulas can be split into two sets:
classical (those that can be used as many times as needed) or
linear (those that are consumed after being used).
Recall that, in a dyadic system for LL, sequents take the form
$
  \Gamma ~;~ \Delta 
  \vdash G $ 
where $G$  is the formula (goal) to be proved (examples in Section \ref{sec:cancer-proofs}), 
 $\Gamma$ is the set of classical formulas and  $\Delta$ is the multiset of linear formulas. 
We store in   $\Gamma$   the formulas representing the rules of the system and in  $\Delta$ the atomic predicates representing the state of the system, namely: 

\noindent \ -   $\bf \underline{\cellF(n,c,f,lm)}$,  denoting  a cell $n$ (a natural number used as an id),
in a compartment $c$ (breast, blood, or bone),
with a phenotype given by a {\em fitness} degree $f \in  \interval{0}{12}$ 
and a list of driver mutations $lm$. The list of driver mutations $lm$ is built up from mutations
$\tgfbeta$, $\epcam$, $\cdfs$, $\cdff$, and/or $\met$,
to which we add $\seeded$, for the cells seeded in the bone.
%
As $\tgfbeta$ is required before any further  mutations,
a list of mutations $[\epcam,\ldots]$ 
will by convention mean $[\tgfbeta,\epcam,\ldots]$;

\noindent \ - $\bf \underline{\apopF(n)}$, representing the fact that the cell $n$ has gone to apoptosis;

\noindent \ - and   ${\bf \underline{\timeF(t)}}$, stating that the current time-unit is $t$. 

Each  rule of the biological system is associated with
a delay (see the terms of the form $d_i$ in Fig. \ref{fig:rules}), which depends on the fitness parameter.
Fitness parameters decrease marginally  with passenger mutations
and increase drastically with driver mutations. 
A typical rule in our model is then as follows: 

\noindent
\resizebox{\textwidth}{!}{
\fbox{$~~  
\ruleF(br_{e0.1})  \eqdef   
   \forall t, n. \cn{T}(t) \otimes \cn{C}(n,breast,1,[\epcam])
   \limp \cn{T}(t+d_{20}(1)) \otimes \cn{C}(n,breast,0,[\epcam])
$}
}
\\[2pt]%
This rule describes a cell acquiring passenger mutations.
Its fitness is decreased by one in a time-delay $d_{20}(1)$ 
and its driver mutations remain unchanged.

Most of the rules in our model are parametric on the fitness degree.
Hence, a rule of the form:

\noindent
\resizebox{\textwidth}{!}{
\fbox{$ 
\ruleF(br_{t1})   \eqdef  \forall t,n.~  
\cn{T}(t) \otimes \cn{C}(n, breast,f,[\tgfbeta]) \limp
\cn{T}(t + d_{11}(f)) \otimes \apopF(n), ~ f \in 0..2$}}
\\[2pt]
represents, in fact, three rules (one for each value of $f\in 0..2$). 
This family of rules describes three cases of apoptosis.
In this particular example, any cell located in the breast, with fitness degree $0$, $1$, or $2$
and list of mutations $[\tgfbeta]$ may go to apoptosis
and the time needed for such a transition is $d_{11}(f)$.
Note that  $d(\cdot)$ is a function that depends on $f$.  
If such $d(\cdot)$ does not depend on $f$,
we shall simply write $d$ instead of $d()$. 

A typical rule describing a cell acquiring a driver mutation is:

\noindent
\resizebox{\textwidth}{!}{
\fbox{$~~
\ruleF(br_{t2.1})   \eqdef    \forall t,n.~ 
\cn{T}(t) \otimes \cn{C}(n, breast,1,[\tgfbeta]) \limp 
\cn{T}(t+d_{12}) \otimes \cn{C}(n, breast,2,[\epcam])
$}}
\\[2pt]
This rule says that a cell in the breast with a fitness degree $f =1$ may acquire a new mutation ($\epcam$),
which increases its fitness by $1$.

Another kind of rule describes a cell moving from one compartment to the next.
The following rule describes an intravasating CTC:

\noindent
\fbox{
\resizebox{.97\textwidth}{!}{
$
\ruleF(br_{e2})  \eqdef  \forall t,n.~
   \cn{T}(t) \otimes \cn{C}(n,breast,f,[\epcam])
   \limp \cn{T}(t+d_{22}(f)) \otimes \cn{C}(n,blood,f+1,[\epcam]), ~ f \in 1..3
$}
}

\begin{figure}[t]
\resizebox{.999\textwidth}{!}{
$
\begin{array}{lll}
\hline \multicolumn{3}{c}{\mbox{In the breast}} \\\hline
\ruleF(br_{0}) & \eqdef &   
   \cn{T}(t) \otimes \cn{C}(n,breast,1,[])
\limp \cn{T}(t+d_{00}) \otimes \cn{C}(n,breast,0,[]) 
\\
\ruleF(br_{1}) & \eqdef &  
\cn{T}(t) \otimes \cn{C}(n,breast,f,[])
\limp \cn{T}(t+d_{01}(f))  \otimes \apopF(n) ~ f \in \interval{0}{1}
\\
\ruleF(br_{2}) & \eqdef &   
   \cn{T}(t) \otimes \cn{C}(n,breast,1,[])
\limp \cn{T}(t+d_{02}) \otimes \cn{C}(n,breast,1,[\tgfbeta])
\\
\ruleF(br_{t0}) & \eqdef &   
   \cn{T}(t) \otimes \cn{C}(n,breast,1,[\tgfbeta])
   \limp \cn{T}(t+d_{10}) \otimes \cn{C}(n,breast,0,[\tgfbeta])
\\
\ruleF(br_{t0r}) & \eqdef &   
   \cn{T}(t) \otimes \cn{C}(n,breast,1,[\tgfbeta])
   \limp \cn{T}(t+d_{10r}) \otimes \cn{C}(n,breast,2,[\tgfbeta])
\\
\ruleF(br_{t1})  &\eqdef&   
   \cn{T}(t) \otimes \cn{C}(n, breast,f,[\tgfbeta]) \limp
   \cn{T}(t + d_{11}(f))   \otimes \apopF(n),  ~ f \in \interval{0}{2}
\\
\ruleF(br_{t2}) & \eqdef &   
   \cn{T}(t) \otimes \cn{C}(n, breast,f,[\tgfbeta]) \limp 
   \cn{T}(t+d_{12}) \otimes \cn{C}(n, breast,f+1,[\epcam]) ~ f \in \interval{1}{2}
\\
\ruleF(br_{e0}) & \eqdef &    
   \cn{T}(t) \otimes \cn{C}(n,breast,f,[\epcam]) \limp
   \cn{T}(t+d_{20}(f)) \otimes \cn{C}(n,breast,f-1,[\epcam]), ~f \in \interval{1}{3}
\\
\ruleF(br_{e0r}) & \eqdef &   
   \cn{T}(t) \otimes \cn{C}(n,breast,f,[\epcam])
   \limp \cn{T}(t+d_{20r}(f)) \otimes \cn{C}(n,breast,f+1,[\epcam]), ~f \in \interval{1}{2}
\\
\ruleF(br_{e1}) & \eqdef &    
   \cn{T}(t) \otimes \cn{C}(n,breast,f,[\epcam])
   \limp \cn{T}(t+d_{21}(f))   \otimes \apopF(n),  f \in  \interval{0}{3}
\\
\ruleF(br_{e2}) & \eqdef &    
   \cn{T}(t) \otimes \cn{C}(n,breast,f,[\epcam])
   \limp \cn{T}(t+d_{22}(f)) \otimes \cn{C}(n,blood,f+1,[\epcam]), f \in \interval{1}{3} 
   \\
\hline \multicolumn{3}{c}{\mbox{In the blood}} \\\hline
\ruleF(bl_{e0}) & \eqdef &   
   \cn{T}(t) \otimes \cn{C}(n,blood,f,[\epcam])  \limp
   \cn{T}(t+d_{30}(f)) \otimes \cn{C}(n,blood,f-1,[\epcam]), f \in \interval{1}{4}
\\
\ruleF(bl_{e0r}) & \eqdef &   
   \cn{T}(t) \otimes \cn{C}(n,blood,f,[\epcam])  \limp
   \cn{T}(t+d_{30r}(f)) \otimes \cn{C}(n,blood,f+1,[\epcam]), f \in \interval{1}{3}
\\
\ruleF(bl_{e1}) & \eqdef &    
   \cn{T}(t) \otimes \cn{C}(n,blood,f,[\epcam])
   \limp \cn{T}(t+d_{31}(f)) \otimes \apopF(n),  f \in \interval{0}{4}
\\
\ruleF(bl_{e2}) & \eqdef &   
   \cn{T}(t) \otimes \cn{C}(n,blood,f,[\epcam]) \limp
   \cn{T}(t+d_{32}(f)) \otimes \cn{C}(n,blood,f+2,[\epcam,\cdfs]), f \in \interval{1}{4}
   \\
\ruleF(bl_{ec0}) & \eqdef &    
   \cn{T}(t) \otimes \cn{C}(n,blood,f,[\epcam,\cdfs]) \limp
   \cn{T}(t+d_{40}(f)) \otimes \cn{C}(n,blood,f-1,[\epcam,\cdfs]),  f \in \interval{1}{6}
\\
\ruleF(bl_{ec1}) & \eqdef &    
   \cn{T}(t) \otimes \cn{C}(n,blood,f,[\epcam,\cdfs])
   \limp \cn{T}(t+d_{41}(f))   \otimes \apopF(n),  f \in \interval{0}{6}
  \\
\ruleF(bl_{ec2}) & \eqdef &    
   \cn{T}(t) \otimes \cn{C}(n,blood,f,[\epcam,\cdfs])  \limp
   \cn{T}(t+d_{42}(f)) \otimes \cn{C}(n,blood,f+2,[\epcdcd]),  f \in \interval{1}{6}
  \\
\ruleF(bl_{ec3}) & \eqdef &     
    \cn{T}(t) \otimes \cn{C}(n,blood,f,[\epcam,\cdfs])   \limp
    \cn{T}(t+d_{43}(f)) \otimes \cn{C}(n,blood,f+2,[\epcdme]),  f \in \interval{1}{6}
   \\
\ruleF(bl_{ecc0}) & \eqdef &    
   \cn{T}(t) \otimes \cn{C}(n,blood,f,[\epcdcd])  \limp \cn{T}(t+d_{50}(f)) \otimes \cn{C}(n,blood,f-1,[\epcdcd]), f \in \interval{1}{6}
\\
   \ruleF(bl_{ecc1}) & \eqdef &    
   \cn{T}(t) \otimes \cn{C}(n,blood,f,[\epcdcd])
   \limp \cn{T}(t+d_{51}(f))  \otimes \apopF(n),  f \in \interval{0}{8}
\\
   \ruleF(bl_{ecc2}) & \eqdef &    
   \cn{T}(t) \otimes \cn{C}(n,blood,f,[\epcdcd])   \limp \cn{T}(t+d_{52}(f)) \otimes \cn{C}(n,blood,f+2,[\epcdcdme]),  f \in \interval{1}{8}
\\
   \ruleF(bl_{ecm0}) & \eqdef &    
   \cn{T}(t) \otimes \cn{C}(n,blood,f,[\epcdme])   \limp \cn{T}(t+d_{60}(f)) \otimes \cn{C}(n,blood,f-1,[\epcdme]),  f \in \interval{1}{8}
\\
   \ruleF(bl_{ecm1}) & \eqdef &    
   \cn{T}(t) \otimes \cn{C}(n,blood,f,[\epcdme])
   \limp \cn{T}(t+d_{61}(f))  \otimes \apopF(n),  f \in \interval{0}{8}
\\
   \ruleF(bl_{ecm2}) & \eqdef &    
   \cn{T}(t) \otimes \cn{C}(n,blood,f,[\epcdme])   \limp \cn{T}(t+d_{62}(f)) \otimes \cn{C}(n,blood,f+2,[\epcdcdme]),  f \in \interval{1}{8}
\\
   \ruleF(bl_{eccm0}) & \eqdef &    
   \cn{T}(t) \otimes \cn{C}(n,blood,f,[\epcdcdme]) 
\limp \cn{T}(t+d_{70}(f)) \otimes \cn{C}(n,blood,f-1,[\epcdcdme]), f \in \interval{1}{10}
\\
   \ruleF(bl_{eccm1}) & \eqdef &    
   \cn{T}(t) \otimes \cn{C}(n,blood,f,[\epcdcdme])
 \limp \cn{T}(t+d_{71}(f))   \otimes \apopF(n),  f \in \interval{0}{10}
\\
   \ruleF(bl_{eccm2}) & \eqdef &    
   \cn{T}(t) \otimes \cn{C}(n,blood,f,[\epcdcdme]) 
  \limp \cn{T}(t+d_{72}) \otimes \cn{C}(n,bone,f+1,[\epcdcdme]),  f \in \interval{1}{10}
\\
  \hline \multicolumn{3}{c}{\mbox{In the bone}} \\\hline
     \ruleF(bo_0) & \eqdef &    
   \cn{T}(t) \otimes \cn{C}(n,bone,f,[\epcdcdme])
 \limp \cn{T}(t+d_{80}(f)) \otimes \cn{C}(n,bone,f-1,[\epcdcdme])
  ,  f \in \interval{1}{11}
\\
   \ruleF(bo_1) & \eqdef &    
   \cn{T}(t) \otimes \cn{C}(n,bone,f,[\epcdcdme])
   \limp \cn{T}(t+d_{81}(f))   \otimes \apopF(n),  f \in \interval{0}{11}
\\
   \ruleF(bo_2) & \eqdef &    
   \cn{T}(t) \otimes \cn{C}(n,bone,f,[\epcdcdme])
  \limp \cn{T}(t+d_{82}(f)) \otimes \cn{C}(n,bone,f+1,[\epcdcdme,\seeded]),  f \in \interval{1}{11}
\end{array}
$
}
\caption{\scriptsize  
Complete set of rules. 
Variables $t$ and $n$ are universally quantified.
$\epcdcdme$, $\epcdcd$ and $\epcdme$ are  shorthand, respectively, 
for the list of mutations  $[\epcam, \cdfs,\cdff,\met]$, $[\epcam, \cdfs,\cdff]$ and  $[\epcam, \cdfs,\met]$. 
\label{fig:rules}}
\end{figure}

Finally, a last kind of rule describes a DNA repair of passenger mutations:

\noindent
\resizebox{\textwidth}{!}{
\fbox{$~~  
\ruleF(br_{e0r})  \eqdef   
   \forall t, n. \cn{T}(t) \otimes \cn{C}(n,breast,f,[\epcam])
   \limp \cn{T}(t+d_{20r}(f)) \otimes \cn{C}(n,breast,f+1,[\epcam]), ~f \in 1..2
$}
}

The complete set of rules is in Fig. \ref{fig:rules}. 
For the sake of readability, we omit the universal quantification on $t$ and $n$ in the formulas.
We shall use $\systemF$ to denote the set of rules and then, sequents take the form:

{\centering$\scriptsize
  \systemF ~;~ \cn{T}(t), 
  \cn{C}(\cdot),  \cdots,  \cn{C}(\cdot) 
  \vdash G
  $\par
}
\noindent
where $G$ is a property to be proved (Sect. \ref{sec:cancer-proofs}). 
Observe that a cell that has gone to apoptosis cannot evolve any further (as no  rule has a   $\apopF(n)$  on the  left hand side).

We note that our rules are asynchronous: only one rule can be fired at a time.
As in Biocham, we choose an asynchronous semantics,  
which enables a more refined description of the evolution of biological systems than the synchronous approach. 
Finally, we note that the delays depend on the fitness and more accurate values can be found using data. 
DNA mutational processes have been succesfully modeled as compound poisson processes 
(see for instance \cite{ding:2013}). 
Delays could be seen as event waiting times which are well-known measures for 
Poisson processes (see for instance \cite{kingman:2009poisson-processes}). 
In our model, delays are (uninterpreted) logical constants that can be later tuned 
when experimental results are available. 
The proofs presented here remain the same regardless of such values. 

\subsection{Verifying Properties of the Model} \label{sec:cancer-proofs}
 
We present here some properties of our model, aiming at 
testing our rules, but also testing some hypotheses of our model---as these are recent proposals in the literature.
The reader may find more properties formalised in \cite{Despeyroux-Felty-Lio-Olarte:mlcsb18}.
We shall detail some of the proofs here. The others can be found 
in the proof scripts and the  documentation of our 
Coq formalisation (\url{http://subsell.logic.at/bio-CTC/}).

The properties we have proven are of three types: 
\begin{enumerate}
\item{} Reachability Properties. The goal here is to prove the existence of a "trace" of execution in the model;
\item{} The existence of cycles. This type of property can be considered a property of reachability, 
although the proofs are often more difficult in other approaches - this is not the case here;
\item{} "Must" properties. Unlike the previous two cases, this is about proving, 
not that a certain behavior \emph{can} take place, but that it \emph{must}: \emph{may vs must} properties.
"Must" properties require inductive reasoning; at the meta level~\footnote{As breafly said in the introduction, our object systems are biological systems, our specification/modelisation logic is linear logic, while our meta logic, used to write proofs, is CCind, on which the Coq proof assistant is based.}.
\end{enumerate}

\subsubsection{A Reachability Property: an \emph{extravasating} CTC}  \label{sec:reach}
A first property of interest describes an extravasating CTC: 
in our case a CTC that has reached the bone.
Let us consider the following property:
``is it possible for a  CTC, with mutations $[\epcam,\cdfs]$ and fitness $3$, to become an extravasating CTC with fitness $8$? What is the time delay for such a transition?''
Recall that a CTC in the blood is a cell $\cn{C}(n, blood, f, m)$ while
an \emph{extravasating CTC}, in the bone, 
is a cell of the shape $\cn{C}(n, bone, f, [\epcam,\cdfs,\cdff,\met])$.
Our property is formalised as follows: 
\begin{proposition} [Reachability]   
\label{prop:cancer-prop1}
The following sequent is provable:

{\centering $\footnotesize
\begin{array}{lll}
  \systemF~;~ .
  & \vdashseq & \forall n,t.~
                \cn{T}(t) \otimes \cn{C}(n,blood,3,[\epcam,\cdfs]) 
\\
  && \limp \exists d.~
     \cn{T}(t+d) \otimes \cn{C}(n,bone,8,[\epcam,\cdfs,\cdff,\met]) 
   \end{array}
$\par
}
\end{proposition}
In our Coq formalisation, we have implemented several tactics (e.g., \texttt{solveF} 
and \texttt{applyRule} used below) 
to automate the process of proving properties  and make the resulting scripts compact and clear.
This should ease the testing/proving of new hypotheses in our model. 
For instance, the proof of the previous property is as follows ($F$ below is the formula in Property 
\ref{prop:cancer-prop1}):
\begin{lstlisting}
Lemma Property1: forall n t, |-  System ; F
Proof with solveF .  (* solves the "trivial" goals in a focused proof *)
  intros. (* introducing the quantified variables n and t *)
  applyRule (blec2 3). (* application of macro rules -- corollary 2-- *)
  applyRule (blecc2 5).
  applyRule (bleccm2 7).
  eapply tri_dec1 ...   (* decision rule, focusing on the goal *)
  eapply tri_tensor ... (* tensor *)
  eapply tri_ex with (t:= (d72 7) s+ (d52 5) s+ (d42 3) s+ (Cte t)) ... 
       (* existential quantifier *)
  eapply Init1... (* initial rule *)
  eapply Init1... (* initial rule *)
Qed.
\end{lstlisting}

The reader may compare the steps in the script above with the proof (by hand) of Property \ref{prop:cancer-prop1} 
in \cite{Despeyroux-Felty-Lio-Olarte:mlcsb18} (Appendix).

\subsubsection{Existence of Cycles} 
Rules for passenger mutations decrease the fitness of the cell by one, while rules for DNA repair increase the fitness. 
Hence, we may observe  loops and oscillations  in our model.  
This can be exemplified in the following property:
``a cell in the breast, with mutation $[\epcam]$, might have its fitness
oscillating from 1 to 2 and back.'' 

\begin{proposition} [Cycle]  
\label{prop:cancer-prop3}
The following sequents are provable:

\resizebox{.9\textwidth}{!}{
$\begin{array}{lll}
  \systemF~;~ .
  & \vdashseq & \forall n,t.~
                \cn{T}(t) \otimes \cn{C}(n,breast,1,[\epcam]) 
     \limp \exists d.~
     \cn{T}(t+d) \otimes \cn{C}(n,breast,2,[\epcam]) 
   \end{array}
$}
and

\resizebox{.95\textwidth}{!}{
$\begin{array}{lll}
  \systemF~;~ .
  & \vdashseq & \forall n,t.~
                \cn{T}(t) \otimes \cn{C}(n,breast,2,[\epcam]) 
     \limp \exists d.~
     \cn{T}(t+d) \otimes \cn{C}(n,breast,1,[\epcam]) 
   \end{array}
$}
\end{proposition}

\subsubsection{A "Must" Property} \label{sec:meta-prop} 
In a first experiment on using LL for biology on the computer
\cite{DeMaria-Despeyroux-Felty:14-fmmb},
we defined the set of biological rules as an inductive type in CIC, and proved
some of their properties by induction on the set of fireable rules.
Here, we choose a different approach.  We have defined the biological rules by formulas in LL,
and we use focusing, along with adequacy, to look for the  fireable rules at a given state. 
Properties of type "must", whose proofs need meta-reasoning, will be formalised at the level of derivations, 
as illustrated here.

The following property states one of the key properties of our model:
``any cell in the blood,  with mutations including $\cdfs$, 
has four possible evolutions:
\begin{enumerate}
\item{} acquiring passenger mutations: 
  its fitness decreases by one and the driver mutations remain unchanged;
\item{} going to apoptosis;
\item{} acquiring a driver mutation: its fitness increases by two;
\item{} moving to the bone:
  its fitness increases by one and the driver mutations ($[\epcam,\cdfs,\cdff,\met]$) remain unchanged.''
\end{enumerate}

\begin{proposition} [Must$_1$]  
\label{prop:cancer-meta2}
Let
$\Delta$ be a multiset of atoms of the form $\cellF(\cdot)$.
Then, in any derivation of the form 
\[
  \infer=[\ruleF(\cdot)]
          {\systemF ; \Delta, \cn{T}(t),  \cn{C}(n,blood,f,m) \vdash G}
          {\systemF ; \Delta, \cn{T}(t+t_d), St \vdash  G}
\]
with $m$ containing $\cdfs$, 
it must be the case that 
\begin{enumerate}
\item{} either $St = \cn{C}(n,blood,f-1,m)$,
\item{} or $St = \apopF(n)$,
\item or $St = \cn{C}(n,blood,f+2,m')$ with $m'$ being as $m$ plus an additional mutation,
\item{} or $St = \cn{C}(n,bone,f+1,m)$ with $m=[\epcam,\cdfs,\cdff,\met]$.
\end{enumerate}
\end{proposition}

In our Coq formalisation,  the above property can be discharged with  few lines of code: 
\begin{lstlisting}
Lemma Property4: forall  n t c lm , F.
Proof by solveF .
 intros H. (* the first sequent is assumed to be provable *) 
 apply FocusOnlyTheory in H;auto. 
      (* The proof H must start by focusing on one of the rules *)
 destruct H as [R] ;destruct H.
repeat (first [ CaseRule  | DecomposeRule; FindUnification | SolveGoal]) . 
Qed.
\end{lstlisting}
The \texttt{FocusOnlyTheory} lemma says that the proof of the sequent must  start  by focusing on one of the formulas in \texttt{System}. The \texttt{destruct} tactic  simplifies the hypotheses after the use of lemma \texttt{FocusOnlyTheory}. The interesting part is the last line of the script. The \texttt{CaseRule} tactic tests each of the rules of the system. Then, \texttt{DecomposeRule; FindUnification} decomposes (positive-negative phase) the application of the rule. Finally, \texttt{SolveGoal} proves the desired goal after the application of the rule. This is a very general scheme, where 
we do case analysis on all possible rules. Some of them cannot be fired in the current state and then, the proof follows by contradiction. In the rest of the cases, the \texttt{SolveGoal} tactic is able to conclude the goal.


\subsection{Conclusion and Future Perspectives on the Biomedicine Section} \label{sec:biomed-conc} 

Here we have proposed first steps in translating logic into biomedicine.  
In our first experiment,
our goal was to study cancer progression, 
aiming at a better understanding of it, and, in the long term, help in finding, and testing, new targeted drugs, 
a priori much more  efficient than most of the drugs used so far.
This chapter describes the use of Linear Logic in modeling the multi-compartment
role of driver mutations in breast cancer. This work is
innovative but also proof-of-principle. It can clearly be generalised
to other cancer types where driver mutations are known.
It also makes evident the capability of this logical approach to integrate
different types of data as basic as mutations in single cell genetics and contribute towards a diagnosis with higher interpretability than many currently fashionable machine learning methods such as deep learning. 
Note however that building a decision support system for cancer/disease diagnosis and therapy prognosis
would require both automatic proof search and various additional information such as the size of the tumour that we do not address here.

Also note that,
although all the properties considered so far only deal with the evolution of one single cell,
our approach allows us to consider a state with many cells. 
We believe that the chapter and the rich sections in the online 
supplementary material (http://subsell.logic.at/bio-CTC/)
will 
become an important resource for other similar studies.
For example we believe 
work on mathematical models such as \cite{AscolaniOcchipintiLio2015}, 
which include survival data and quantitative results, is complementary to our work.
Logic allows the modeling of the evolution of cells across scales and compartments, 
while ODEs require parameter estimation (qualitative vs quantitative).

While temporal logics  
have been very successful in practice
with efficient, fully automatic, model checking tools, 
these logics do not enjoy standard proof theory.  
In contrast, Linear Logic has a very traditional proof theoretic pedigree:
it is naturally presented as a sequent calculus that enjoys cut-elimination and focusing.
A further advantage of our approach with respect to model checking is that
it provides a unified framework to encode both transition rules  
and (both statements and proofs of)
temporal properties. 
Observe also that we do not need to build the set of states of the transition system
(neither completely nor "on-the-fly").
Last but not least, as we shall see in the next section devoted to neuroscience,
the computational logic approach enables proofs of more general properties, compared to model-checking; 
thanks to the expressiveness of the logic language (here ILL and/or CIC)
and to the powerful associated proof tools (as the Coq system chosen here).
We view model checking  as a useful first step before proofs:
testing the model before trying to prove properties of it. 
The interested reader can find a detailed comparison of the approaches 
in \cite{DeMaria-Despeyroux-Felty:14-fmmb}.
See also \cite{Despeyroux-Olarte-Pimentel:entcs-17} for 
an adequate encoding of temporal logic in Linear Logic. 

We believe that different modeling approaches should be linked, where relevant. 
For our part, we plan, in the near future, 
not, a priori, develop our own interface to a "biology dedicated LL", but 
propose translations/compilations from models written in rule-based
languages such as Biocham and Kappa (presented in other chapters of
the present book) to corresponding models written in LL. 
This compilation seems feasable, at least without taking into account reaction rates, 
or by translating them into delays where appropriate.
This would provide the biologist with a richer environment in which to model his systems
and carry out initial static, and possibly causal, analyses of his models in Biocham or Kappa, 
before proving their (or other) properties of interest in logic.

\section{On the Use of Coq to Model and Verify Neuronal Archetypes}
\label{sec:lf-neuro}

\subsection{\label{sec:Neuro-Intro}Introduction}
In recent years, the exploration of neuronal micro-circuits has become
an emerging question in Neuroscience, notably with respect to
their integration with neurocomputing approaches \cite{Markram2006}. \emph{Archetypes} \cite{DMGRG16HSB} are specific graphs of a few neurons with biologically relevant structure and behavior. These archetypes correspond to elementary and fundamental elements of neuronal information processing. Several archetypes can be coupled to constitute the elementary building blocks of bigger neuronal circuits in charge of specific functions. For instance, locomotive motion and other rhythmic behaviors are controlled by well-known specific neuronal circuits called Central Generator Patterns (CPG) \cite{M87BC}.
These CPG circuits have an oscillating output when the neurons composing them have some specific parameters and when these neurons are connected in a specific way.

To study these circuits,
it is relevant to investigate the dynamic behavior of all the possible archetypes of 2, 3 or more neurons, up to and including the consideration of archetypes of archetypes.
One of the open questions is: are the properties of these archetypes of archetypes simply the conjunction of the individual constituent archetypes properties or something more or less? In other words, does the resulting network satisfy exactly the properties that were already satisfied by the constituent archetypes or are there new properties (or less properties) that are satisfied by this combination?  Another crucial question is: can we understand the computational properties of large groups of neurons simply as the coupling of the properties of individual archetypes, as it is for the alphabet and words, or is there something more again?

The first attempts in the literature to apply \emph{formal methods} from computer science to  model and verify temporal properties of fundamental neuronal archetypes in terms of neuronal information processing can be found in \cite{DMGRG16HSB, DLGRG17CSBIO}. In this work, the authors  take advantage of the synchronous programming language Lustre to implement
six neuronal archetypes and their coupling and to formalize their expected properties.
Then, they exploit some associated model checkers to automatically validate these behaviors. However, model checkers prove properties for some given parameter intervals, and do not handle inputs of arbitrary length. In the work described in this section, we use Coq to prove four important properties of neurons and archetypes.  This work extends \cite{BDF18CSBIO}, primarily by considering archetypes in fuller detail. One of the main advantages of using Coq for the work in this section is the generality of its proofs. Using such a system, we can prove properties about arbitrary values of parameters, such as any length of time, any input sequence, or any number of neurons (parametric verification). We use Coq's general facilities for structural induction and case analysis, as well as Coq's standard libraries that help in reasoning about rational numbers and functions on them.

As far as neuronal networks are concerned, in the literature their modeling is classified into three generations \cite{Maass97,Moisy12}.
First generation models, represented by the McCulloch-Pitts model \cite{MW43BMB}, handle discrete inputs and
outputs and their computational units consist of a set of logic gates with a threshold activation function.
Second generation models, whose most representative example is the multi-layer perceptron \cite{cybenko89}, exploit real valued activation functions.
These networks, whose real-valued outputs represent neuron firing rates, are widely used in the domain of artificial intelligence.
Third generation networks, also called \emph{spiking neural networks} \cite{Moisy12}, are characterized by the relevance of time aspects. Precise
spike firing times are taken into account. Furthermore, they consider not only current input spikes but also past ones (temporal summation).
In \cite{izhikevich04}, spiking  neural networks are classified with respect to their biophysical plausibility, that is, the number of behaviors (i.e., typical responses to an input pattern) they can reproduce. Among these models, the Hodgkin-Huxley model \cite{hodgkin52} is the one able to reproduce most behaviors. However, its simulation process is very expensive even for a few neurons and for a
small amount of time. In this work, we choose to use the leaky integrate and fire (LI\&F) model \cite{lapicque1907}, a computationally efficient
approximation of a single-compartment model, which proves to be amenable to formal verification.
Notice that discrete modeling is well suited for this task because neuronal activity, as with any other recorded physical event, is only recorded at discrete intervals (the recording sampling rate is usually set at a significantly higher resolution than the rate of the system being recorded, so that there is no loss of information). We describe neural networks as weighted directed graphs whose nodes represent neurons and whose edges stand for synaptic connections. At each time unit, all the neurons compute their membrane potential accounting not only for the current input signals, but also for the signals received during a given temporal window. Each neuron can emit a spike when it exceeds a given threshold.

In addition to the papers already cited in this section (\cite{DMGRG16HSB, DLGRG17CSBIO, BDF18CSBIO}), in the literature there are a few attempts at giving formal models for spiking neural networks. In \cite{ciobanu16}, a mapping of spiking neural P systems into timed automata is proposed. In that work, the dynamics of neurons are expressed in terms of evolution rules and durations are given in terms of the number of rules applied. Timed automata are also exploited in \cite{DD18B} to model LI\&F networks. This modeling is substantially different from the one proposed in \cite{ciobanu16} because an explicit notion of duration of activities is given. Such a model is formally validated against some crucial properties defined as temporal logic formulas and is then exploited to find an assignment for the synaptic weights of neural networks so that they can reproduce a given behavior.

The rest of this section is organized as follows. In Section \ref{sec:LIF} we present a discrete version of the LI\&F model. Section \ref{sec:Archetypes} is devoted to the description of the neuronal archetypes we consider. In Section \ref{sec:coq}, we introduce features of Coq important for this work.
In Section \ref{sec:coq-encoding}, we present the Coq model for neural networks, which includes definitions of neurons, operations on them, and the combination of neurons into archetypes.  In Section \ref{sec:EncodingCoq}, we present and discuss four important properties, starting with properties of single neurons and the relation between the input and output, and moving toward more complex properties that express their interactions and behaviors as a system.
Proofs can be found in \cite{BDF18CSBIO}. Finally, in Section \ref{sec:Neuro-Conclusion} we conclude and discuss future work. The accompanying Coq code can be found at:\\ \url{http://www.site.uottawa.ca/~afelty/coq-archetypes/}.

\subsection{\label{sec:LIF}Discrete Leaky Integrate and Fire Model}
In this section, we introduce a discrete (Boolean) version of LI\&F modeling. We first
present the basic biological knowledge associated to the phenomena we
model, and then we provide details of the model.

When a neuron receives a signal at one of its synaptic connections, it produces
an excitatory or an inhibitory \textit{post-synaptic potential} (PSP) caused by
the opening of selective ion channels according to the nature of the
post-synaptic receptor. An activation leads to an inflow of cations in the cell; an inhibition
leads to an inflow of anions in the cell. This local flow of ions influences the
electrochemical potential difference on both sides of the plasma membrane and
locally depolarizes (excitation) or hyper-polarizes (inhibition) the neuron
membrane. Such polarization is transmitted, step by step, to the rest of the
membrane and thus influences the potential difference on both sides of the
membrane at the whole cell level. The potential difference is called
\textit{membrane potential}.
In general, a PSP alone is not enough for
the membrane potential of the receiving neuron to exceed its
\textit{depolarization threshold}, and thus to emit an
\textit{action potential} at its axon to transmit the signal to other neurons.

However, two phenomena allow the cell to exceed its depolarization threshold:
the \textit{spatial summation} and the \textit{temporal summation}
\cite{doi:10.1086/504005}. Spatial summation is the sum of all the different
PSPs produced at a given time at different areas of the membrane. Temporal
summation is the sum of all the different PSPs produced between two
instants that are considered ``close enough.''
These summations are possible thanks to a property of the membrane,
which allows it to
behave like a capacitor and locally store some electrical loads (\textit{capacitive property}). 

The neuron membrane, due to the presence of leakage channels, is not a
perfect conductor and can be compared to a resistor inside an electrical
circuit. Thus, the range of the PSPs decreases with
time and space (\emph{resistivity} of the membrane).

A LI\&F neuron network is represented with a weighted directed graph where each
node stands for a neuron soma and each edge stands for a synaptic connection
between two neurons. The associated weight for each edge is an indicator of the
weight of the connection of the receiving neuron: a positive (resp. negative) weight is an
activation (resp. inhibition). 

The depolarization threshold of each neuron is modeled via the \textit{firing
threshold } $\tau$, which is a numerical value; it is the value that the neuron membrane
potential $p$ must exceed at a given time $t$ in order to emit an action potential, or
\textit{spike}, at the time $t+1$.

The membrane resistivity is modeled as a numerical coefficient called the
\textit{leak factor}  $r$, which allows the range of a PSP to decrease
over time.

Spatial summation is implicitly taken into account. In our model, a neuron $u$
is connected to another neuron $v$ via a single synaptic connection of weight
$w_{uv}$. This connection represents the entirety of the shared connections
between $u$ and $v$. Spatial summation is also more explicitly taken into account with the fact that,
for each instant, the neuron sums each signal received from each input neuron.
%
To take the temporal summation into account, we add both the past and
present PSPs when computing the membrane potential. As the value of a
PSP gets older over time, it has less impact on the calculation of the
current membrane potential.  This decrease is taken into account by
repeatedly decreasing the value by $r$. As a consequence, old PSPs can
be neglected (their value is very small) and the effect of past PSPs
is restricted to a given time interval.  In particular,
we introduce a \textit{sliding integration window} of length $\sigma$ for each neuron.
This allows us to obtain finite sets of states, and thus to easily apply formal techniques.

More formally, the following definition can be given:

\begin{definition}[Boolean Spiking Integrate and Fire Neural Network]
			\label{def.snn}
			\emph{A spiking Boolean integrate and fire neural network} is a tuple $(V,\, E,\, w)$, where: \begin{itemize}
				\item $V$ is a set of Boolean spiking integrate and fire neurons,
				\item $E \subseteq V \times V$ are synapses,
				\item $w: E \rightarrow \mathbb{Q} \cap [-1,1]$ is the synapse weight function associating to each synapse $(u,\, v)$ a weight $w_{uv}$.
			\end{itemize}
            \emph{A spiking Boolean integrate and fire neuron} is a tuple $(\tau, r, p, y)$, where:
            \begin{itemize}
				\item $\tau \in \mathbb{N}$ is the \emph{firing threshold},
                \item $r \in \mathbb{Q} \cap [0, 1]$ is the \emph{leak factor},

                \item $p: \mathbb{N} \rightarrow \mathbb{Q}^+_0$ is the [membrane] \emph{potential} function defined as

                \[
p(t) =\left\{ \begin{array}{c}
\sum_{i=1}^{m} w_i \cdot x_i(t),\quad if\,p(t-1) \geqslant \tau\\
\sum_{i=1}^{m} w_i \cdot x_i(t) + r \cdot p(t - 1),\quad otherwise
\end{array}\right.
\]
		where $p(0)=0$, $m$ is the number of inputs of the neuron, $w_i$ is the weight of the synapse connecting the $i^{th}$ input neuron to the current neuron, and  $x_i(t) \in \{0, 1\}$ is the signal received at the time $t$ by the neuron through its $i^{th}$ input synapse (observe that, after the potential exceeds its threshold, it is reset to $0$),
                \item $y: \mathbb{N} \rightarrow \{0, 1\}$ is the neuron output function,  defined as
                \begin{equation*}
			y(t) = \begin{cases}
				1 & \mathrm{if}\  p(t-1) \geqslant \tau \\
				0 & \mathrm{otherwise.}
			\end{cases}
		\end{equation*}
			\end{itemize}	
		\end{definition}
The development of the recursive equation for the membrane potential function and the introduction of a sliding integration window of length $\sigma$
leads to the following equation for $p$ (when $p(t-1) < \tau$):
$p(t) =\sum_{e=0}^{\sigma}r^{e}\sum_{i=1}^{m} w_i \cdot x_{i} (t-e)$, where $e$ represents the time elapsed up until the current time.

\subsection{\label{sec:Archetypes}The Basic Archetypes}
\begin{figure}[ptbh]
\begin{center}
\includegraphics[scale=.55]{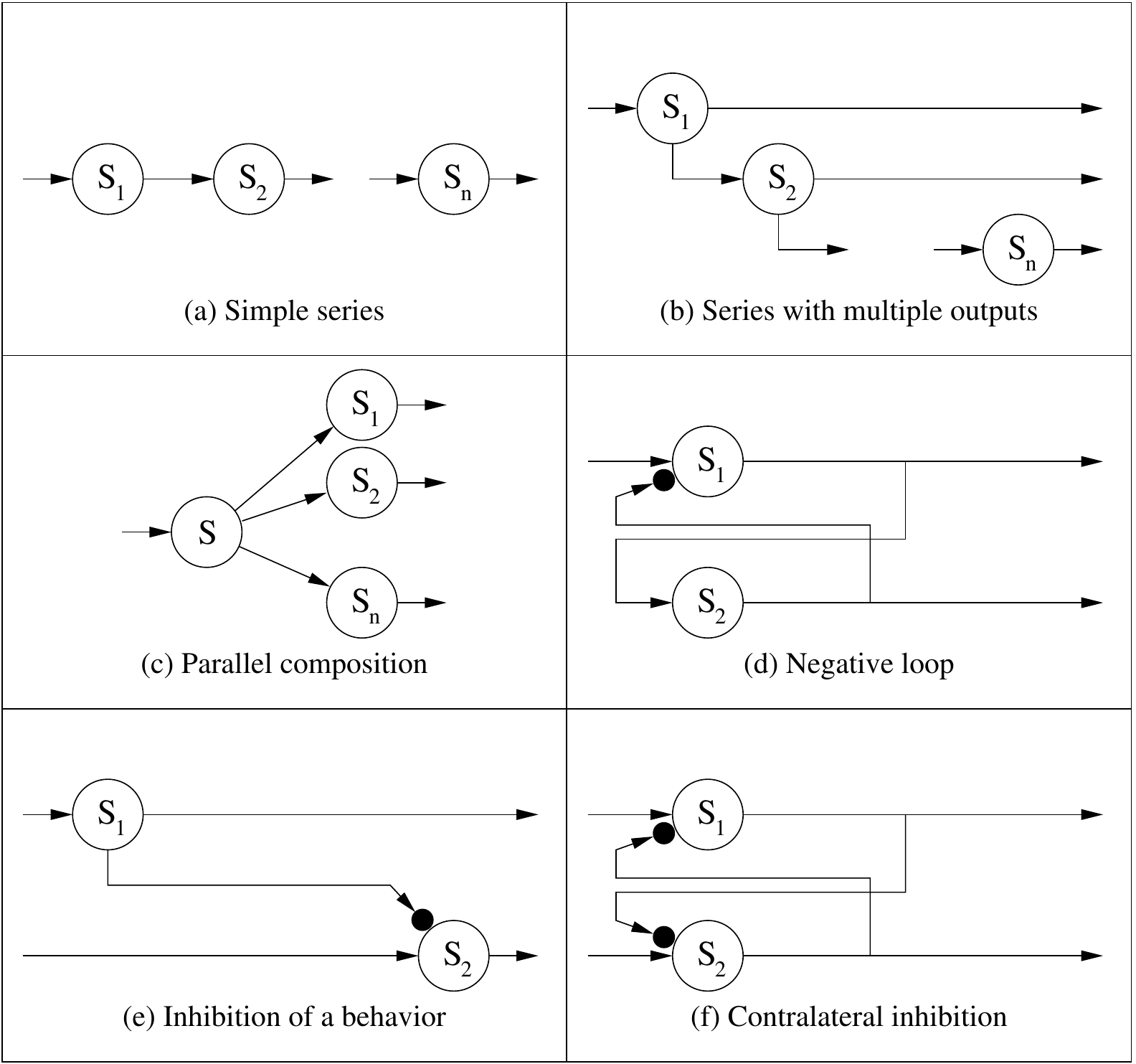}
\end{center}
\caption{The basic neuronal archetypes.}
\label{archetypes}
\end{figure}

The six basic archetypes we study are as follows (see Fig. \ref{archetypes}).
These archetypes can be coupled to constitute the elementary building blocks of bigger neuronal circuits.
\begin{itemize}
\item {\textbf{Simple series}} is a sequence of neurons where each
element of the chain receives as input the output of the preceding
one. 
\item {\textbf{Series with multiple outputs}} is a series where, at
each time unit, we are interested in knowing the outputs of all the
neurons (i.e., the output of all the neurons together constitutes the
output of the archetype as a whole).
\item {\textbf{Parallel composition}} is a set of neurons receiving
as input the output of a given neuron.
\item {\textbf{Negative loop}} is a loop consisting of two neurons:
the first neuron activates the second one while the latter inhibits
the former.
\item {\textbf{Inhibition of a behavior}} consists of two neurons, the first
one inhibiting the second one.
\item {\textbf{Contralateral inhibition}} consists of two or more neurons,
each one inhibiting the other ones.
\end{itemize}

\subsection{\label{sec:coq}Modelling in Coq}
In this section, we present the basic elements of Coq that we use to represent our model.
We encode the model and properties directly in the logic implemented in Coq, which is the Calculus of Inductive Constructions.  We use several basic types, data structures, and properties from Coq's libraries, and we do not add any axioms. Expressions in Coq include a functional programming language. It is a typed language, which means that every Coq expression has a type. For instance, \lstinline{X:nat} expresses that variable \lstinline{X} is in the domain of natural numbers. The types used in our model include \lstinline{nat}, \lstinline{Q}, and \lstinline{list} which denote natural numbers, rational numbers, and list of elements respectively. These types are found in Coq's standard libraries. All elements in a list have the same type. For instance, \lstinline{L:list nat} means that \lstinline{L} is a list of natural numbers. A list can be empty, which is written \lstinline{[]} or \lstinline{nil} in Coq. Functions are a basic element of any functional programming language. The general form of a function in Coq is shown below.
\begin{lstlisting}[escapeinside={/*}{*/}]
Definition/Fixpoint /*Function\_Name*/
/*(Input$_1$: Type of Input$_1$) $\dots$ (Input$_n$: Type of Input$_n$): Output Type*/ := /*Body of the function*/.
\end{lstlisting}
\lstinline{Definition} and \lstinline{Fixpoint} are Coq keywords for defining non-recursive and recursive functions, respectively. After either one of these keywords comes the name that a programmer gives to the function. Following the function name are the input arguments and their types. If two or more inputs have the same type, they can be grouped. For example, \lstinline{(X Y Z: Q)} means all variables \lstinline{X}, \lstinline{Y}, and \lstinline{Z} are rational numbers. Following the inputs is a colon, followed by the output type of the function. Finally, the body of the function is a Coq expression representing a program, followed by a dot.

Pattern matching is a useful feature in Coq, used for case analysis. This feature is used, for instance, for distinguishing between base cases and recursive cases in recursive functions. For example, it can distinguish between empty and nonempty lists. The pattern for a non-empty list shows the first element of the list, which is called the \emph{head}, followed by a double colon, followed by the rest of the list, which is called the \emph{tail}. The tail of a list itself is a list of elements of the same type as the type of the head. For example, let $L$ be the list \lstinline{(6::3::8::nil)} containing three natural numbers.  An alternate notation for Coq lists allows $L$ to be written as \lstinline{[6;3;8]} where the head is \lstinline{6} and the tail is \lstinline{[3;8]}. Thus, the general pattern for non-empty lists often used in Coq recursive functions has the form \lstinline{(h::t)}. Another example of a Coq data type is the natural numbers. A natural number is either \lstinline{0} or the successor of another natural number, written \lstinline{(S n)}, where \lstinline{n} is a natural number. For example, \lstinline{1} is represented as \lstinline{(S 0)}, \lstinline{2} as \lstinline{(S (S 0))}, etc. In the code below, some patterns for lists and natural numbers are shown using Coq's \lstinline{match$\dots$with$\dots$end} pattern matching construct.
\begin{lstlisting}[escapeinside={/*}{*/}]
match X with
 | 0 => /*calculate something when */ X /*=*/ 0
 | S n => /*calculate something when */ X /*is the successor of*/ 0
end

match L with
 | [] => /*calculate something when*/ L /*is an empty list*/
 | h::t => /*calculate something when*/ L /*has head*/ h /*followed by tail*/ t
end
\end{lstlisting}
In addition to the data types that are defined in Coq's libraries, new data types can be defined. One way to do so is using records. Records can have different fields with different types. For example, we can define a record that has 3 fields \lstinline{Fieldnat}, \lstinline{FieldQ}, and \lstinline{ListField}, which have types natural number, rational number, and list of natural numbers, respectively. The code below shows the Coq syntax for the definition of this record with one additional field called \lstinline{CR}.
\begin{lstlisting}
Record Sample_Record := MakeSample {
  Fieldnat: nat;
  FieldQ: Q;
  ListField: list nat;
  CR: Fieldnat > 7 }.

S: Sample_Record
\end{lstlisting}
Fields in Coq can represent conditions on other fields. For example, field \lstinline{CR} in the above code is a condition on the \lstinline{Fieldnat} field stating that it must be greater than 7. After defining a record type, it is a type like any other type, and so for example, we can have variables with the new record type. Variable \lstinline{S} shown with type \lstinline{Sample_Record} is an example.

\subsection{\label{sec:coq-encoding}Encoding Neurons and Archetypes in Coq}
We illustrate our encoding of neural networks in Coq by beginning with the code below.
\begin{lstlisting}
Record Neuron := MakeNeuron {
  Output:list nat;
  Weights:list Q;
  Leak_Factor:Q;
  Tau:Q;
  Current:Q;
  Output_Bin: Bin_List Output;
  LeakRange: Qle_bool 0 Leak_Factor = true /\
             Qle_bool Leak_Factor 1 = true;
  PosTau: Qlt_bool 0 Tau = true;
  WRange: WeightInRange Weights = true }.

Fixpoint potential (Weights: list Q) (Inputs: list nat): Q :=
  match Weights, Inputs with
  | nil, _ => 0
  | _, nil => 0
  | h1::t1, h2::t2 => if (beq_nat h2 0%nat)
                      then (potential t1 t2)
                      else (potential t1 t2) + h1
  end.
\end{lstlisting}
We use Coq's record structure to define a neuron.  This record includes five fields with their types, and four fields which represent constraints that the first five fields must satisfy according to the LI\&F model mentioned in Section \ref{sec:LIF}. The types include natural numbers, rational numbers, and lists.  In particular, a neuron's output \lstinline{(Output)} is represented as a list of natural numbers, with one entry for each time step.  The weights attached to the inputs of the neuron \lstinline{(Weights)} are stored in a list of rational numbers, one for each input in some designated order.  The leak factor \lstinline{(Leak_Factor)}, the firing threshold \lstinline{(Tau)}, and the most recent neuron membrane potential \lstinline{(Current)} are rational numbers.  With respect to the four conditions, for example, consider \lstinline{PosTau}, which expresses that \lstinline{Tau} must be positive. \lstinline{Qle_bool} and other arithmetic operators come from Coq's rational number library. The other three state, respectively, that \lstinline{Output} contains only 0s and 1s (it is a binary list), \lstinline{Leak_Factor} is between 0 and 1 inclusive, and each input weight is in the range of [-1, 1].  We omit the definitions of \lstinline{Bin_List} and \lstinline{WeightInRange} used in these statements.

Given a neuron \lstinline{N}, we write \lstinline{(Output N)} to denote its first field, and similarly for the others.  To create a new neuron with values \lstinline{O, W, L, T,} and \lstinline{C} of the appropriate types, and proofs \lstinline{P1,$\dots$, P4} of the four constraints, we write \lstinline{(MakeNeuron O W L T C P1 P2 P3 P4)}.

The next definition in the above code implements the weighted sum of the inputs of a neuron, which is an important part of the calculation of the potential function (see Definition \ref{def.snn}). In this recursive function, there are two arguments: \lstinline{Weights} representing $w_1,\dots,w_m$ and \lstinline{Inputs} representing $x_1,\dots,x_m$.  The function returns an element of type \lstinline{Q}. Its definition uses pattern matching on both inputs simultaneously.  The body of the definition uses Booleans, the \lstinline{if} statement, and the equality operator on natural numbers \lstinline{(beq_nat)}, all from Coq's standard library. Natural numbers, such as \lstinline{0\%nat} above are marked with their type to distinguish them from rational numbers, whose types are omitted.
 Although, we always call the \lstinline{potential} function with two lists of equal length, Coq requires functions to be total; when two lists do not have equal length, we return a ``default'' value of 0. Also, when we call this function, \lstinline{Inputs}, which is the second argument of the function, will always be a binary list (contains only the natural numbers 0 and 1). Thus, when the head of the list \lstinline{h2} is 0, we do not need to add anything to the final sum because anything multiplied by 0 is 0. In this case, we just call the function recursively on the remaining weights and inputs \lstinline{t1} and \lstinline{t2}, respectively. On the other hand, when \lstinline{h2} is 1, we need to add \lstinline{h1}, the head of \lstinline{Weights} to the final sum, which again is the recursive call on \lstinline{t1} and \lstinline{t2}.

The following code shows the \lstinline{NextPotential} function, which implements $p(t)$ from Definition \ref{def.snn}.
\begin{lstlisting}
Definition NextPotential (N: Neuron) (Inputs: list nat): Q :=
 if (Qle_bool (Tau N) (Current N))
 then (potential (Weights N) Inputs)
 else (potential (Weights N) Inputs) +
      (Leak_Factor N) * (Current N).
\end{lstlisting}
Recall that \lstinline{(Current N)} is the most recent potential value of the neuron which is $p(t-1)$ in Definition \ref{def.snn}. \lstinline{(Qle_bool (Tau N) (Current N))} represents $\tau \leq p(t-1)$ and we use the potential function defined before for the part calculating the weighted sum of the neuron inputs. Finally, the last line implements $r \cdot p(t-1)$.

The following code contains two definitions.
\begin{lstlisting}
Definition NextOutput (N: Neuron) (Inputs: list nat): nat :=
  if (Qle_bool (Tau N) (NextPotential N Inputs))
  then 1%nat
  else 0%nat

Definition NextNeuron (N: Neuron) (Inputs: list nat): Neuron :=
  MakeNeuron
    ((NextOutput N Inputs)::(Output N))
    (Weights N)
    (Leak_Factor N)
    (Tau N)
    (NextPotential N Inputs)
    (NextOutput_Bin_List N Inputs (Output_Bin N))
    (LeakRange N)
    (PosTau N)
    (WRange N).
\end{lstlisting}
The first definition computes the next output of the neuron, which is $y(t)$ in Definition \ref{def.snn}. Recall that \lstinline{(NextPotential N Inputs)} computes $p(t)$. Thus, the expression \lstinline{(Qle_bool (Tau N) (NextPotential N Inputs))} expresses the condition $\tau \leq p(t)$.

 In our model, the state of a neuron is represented by the \lstinline{Output} and \lstinline{Current} fields. The \lstinline{Output} field of a neuron in the initial state is \lstinline{[0\%nat]}, which denotes a list of length 1 containing only 0.  The \lstinline{Current} field represents the initial potential, which is set to 0.  A neuron changes state by processing input.  After processing a list of $n$ inputs, the \lstinline{Output} field will be a list of length $n+1 $ containing 0's and 1's, and the \lstinline{Current} field will be set to the value of the potential after processing these $n$ inputs. State change occurs by applying the \lstinline{NextNeuron} function of the above code to a neuron and a list of inputs.  As it is typical in functional programming, we represent a neuron at its later state by creating a new record with the new values for \lstinline{Output} and \lstinline{Current} and other values directly copied over. We store the values in the \lstinline{Output} field in reverse order, which simplifies proofs by induction over lists, which we use regularly in our Coq proofs. Thus, the most recent output of the neuron is at the head of the list. We can see this in the above code, where the new value of the output is \lstinline{((NextOutput N Inputs)::(Output N))}.  The next output of the neuron is at the head, followed by the previous outputs. \lstinline{(NextPotential N Inputs)} is the new value for \lstinline{(Current N)}. Recall that \lstinline{(Current N)} is the most recent value of potential value of the neuron or $p(t-1)$. So, for calculating the next potential value of the neuron or $p(t)$, the \lstinline{NextPotential} function is called.

Following the new values for each field of the neuron, we have proofs of the four constraints.  The first requires a lemma \lstinline{NextOutput_Bin_List} (statement omitted) which allows us to prove that the new longer list is still a binary list.  Proofs of the other three constraints are carried over exactly from the original neuron, since they are about components of the neuron that do not change.

To reinitialize a neuron to the initial state, the \lstinline{ResetNeuron} function is used.
\begin{lstlisting}
Definition ResetNeuron (N: Neuron): Neuron := MakeNeuron
  ([0%nat])
  (Weights N)
  (Leak_Factor N)
  (Tau N)
  (0)
  (Reset_Output)
  (LeakRange N)
  (PosTau N)
  (WRange N).
\end{lstlisting}
This function takes any \lstinline{Neuron} as input, and returns a new one, with the \lstinline{Output}, \lstinline{Current}, and \lstinline{Output_Bin} fields reset, while keeping the others. The \lstinline{Reset_Output} property is a simple lemma stating that \lstinline{[0\%nat]} satisfies the \lstinline{Bin_List} property.

So far, we have discussed the encoding and processing of single neurons in isolation, which take in inputs and produce outputs.
We next consider archetypes.
In general, our approach is to encode the particular structure of each archetype as a Coq record. Using a record for each archetype facilitates stating and proving properties about them. Recall that archetypes are functional structures of neural networks. Defining them in this abstract way helps us to present their basic functions.
To illustrate this approach, we introduce now the encoding of one archetype, the simple series in Figure \ref{archetypes}(a).
As shown in Figure \ref{archetypes}(a), a simple series consists of a list of single input neurons. The first neuron receives the input of the archetype and sends its output to the second neuron. Starting from the second neuron each neuron receives its input from the previous neuron and sends its output as the input of the next neuron in the series. The last neuron produces the output of the series.
The \lstinline{NeuronSeries} record type defined below represents this structure in Coq.

\begin{lstlisting}
Record NeuronSeries {Input: list nat} := MakeNeuronSeries
{
  NeuronList: list Neuron;
  NSOutput: list nat;
  AllSingle: forall (N:Neuron),
    In N NeuronList -> (beq_nat (length (Weights N)) 1%nat) = true;
  SeriesOutput: NSOutput = (SeriesNetworkOutput Input NeuronList);
}.

\end{lstlisting}

Records can have input parameters, similar to functions in Coq, and here the list of inputs to the simple series is \lstinline{Input:~list nat}. Thus \lstinline{NeuronSeries} is actually a function from a list of natural numbers to a record.
Curly brackets around input arguments is Coq notation for \emph{implicit} arguments, which are arguments that can be omitted from expressions as long as Coq can figure out the missing information. Its use here allows us to write more readable Coq code.
\lstinline{NeuronList} is a field in the record representing the list of neurons in the simple series. The first element in this list is the first neuron in the series, etc. \lstinline{NSOutput} represents the list of outputs of the series. In other words, it is the output list of the last neuron in the series, which is also the last neuron of \lstinline{NeuronList}. There are also two constraints for this archetype. \lstinline{AllSingle} expresses that all neurons in the series are single input neurons. The functions \lstinline{In} and \lstinline{length} are defined in Coq's list library and define list membership and size of a list, respectively. \lstinline{SeriesOutput} expresses that the output of the series is equal to the output of the function \lstinline{SeriesNetworkOutput}. This function takes the input of the series and list of neurons in the series and produces the output of the series. We leave out its definition and just note here that it expresses the details of the input/output connections between the elements of \lstinline{NeuronList}, and in the degenerate case when \lstinline{NeuronList} is empty, \lstinline{NSOutput} is set to the input.

\subsection{\label{sec:EncodingCoq}Properties of Neurons and Archetypes in Coq}
As mentioned earlier, we state four basic properties of the LI\&F model of neurons in this section. All of them have been fully verified in Coq. We start in the next subsection with a property about a simple neuron, which has only one input. We refer to this neuron as a single-input neuron.

In all of the statements of the properties, we omit the assumption that the input sequence of the neuron is a binary list and contains only 0s and 1s. It is, of course, included in the Coq code. We use several other conventions to enhance readability when stating the property. For example, we state our property using pretty-printed Coq syntax, with some abbreviations for our own definitions.  For instance, we use mathematical fonts and conventions for Coq text, e.g., \lstinline{(Output N)} is written $Output(N)$,  \lstinline{(Tau N)} is written $\tau(N)$, \lstinline{(Weights N)} is written $w(N)$, \lstinline{(Leak_Factor N)} is written $r(N)$, and \lstinline{(Current N)} is written $p(N)$. In addition, if $w(N)$ is a list of the form $[w_1;\dots; w_n]$ for some $n\geq0$, for $i=1,\dots,n$, we often write $w_i(N)$ to denote $w_i$.  Also, we use notation and operators from the Coq standard library for lists.  For instance, $length$ and $+$ are list operators; the former is for finding the number of elements in the list and the latter is the notation we will use here for list concatenation.
In addition, although for a neuron $N$, the list $Output(N)$ is encoded in reverse order, in our Coq model, when presenting properties, we use forward order.

\subsubsection{The Delayer Effect for a Single-Input Neuron}
The first property is called the \emph{delayer effect} property. It concerns a single neuron, which has only one input. Recall that a neuron is in an inactive state initially, which means the output of a neuron at time 0 is 0. When a neuron has only one input, and the weight of that input is greater than or equal to its activation threshold, then the neuron transfers the input sequence to the output without any change (except for a "delay" of length 1). For instance, if a single input neuron receives 0100110101 as its input sequence, it will produce 00100110101 as output. Neurons that have this property are mainly just transferring signals. Humans have some of this type of neuron in their auditory system. This property is expressed as Theorem \ref{PropCoq}.

\begin{theorem}
\label{PropCoq} \textbf{[Delayer effect for a single-input neuron]}

$$\forall (N:neuron)  (input:list \; nat),$$
$$ length(w(N))=1 \wedge w_1(N)\geq \tau(N) \rightarrow $$
$$Output(N')=[0]+input$$
\end{theorem}
In the above statement, $N'$ denotes the neuron obtained by initializing $N$ and then processing the input (using $ResetNeuron$ and repeated applications of $NextNeuron$). We use this convention in stating all of our properties.  Note that in Definition \ref{def.snn}, $p$ is a function of time.  Time in our Coq model is encoded as the position in the output list. If $Output(N)$ has length $t$, then $p(N)$ stores $p(t-1)$ from Definition \ref{def.snn}. If we then apply $NextNeuron$ to $N$ and the next input obtaining $N'$, then $Output(N')$ has length $t+1$ and $p(N')$ stores the value $p(t)$ from Definition \ref{def.snn}.
The theorem is proved by induction on the length of the input sequence.

\subsubsection{The Filter Effect for a Single Neuron}
The next property we consider is also about single-input neurons.
It is defined with respect to a given integer $n$, where $n$ is less than or equal to $\sigma$, the length of the integration window introduced in Section \ref{sec:LIF}.  When a neuron has
only one input, and the weight of that input is less than its
activation threshold, the neuron passes on the value 1 once as output
for every $n$ consecutive occurrences of 1s in the
input. For each such $n$ occurrences of 1s in the input, all 1s are
replaced by 0 except for the last one. The other 1s are ``filtered
out.''  The neuron thus only ``passes'' one signal out of $n$, it
behaves as a $1/n$ filter.  As a consequence, there are never two consecutive 1s in the output sequence.  This consequence is called the filter effect. Most neurons in a human body have the filter effect because their input weight is less than their activation threshold. Normally, more than one input is needed to activate a human neuron. In biology, this property is often called the integrator effect.

\begin{theorem}
\label{Prop2} \textbf{[Filter effect for a single-input neuron]}
$$\forall (N:neuron)  (input:list \; nat),$$
$$ length(w(N))=1 \wedge w_1(N)< \tau(N) \rightarrow 11 \notin Output(N')$$
\end{theorem}

Note that in the statement above, $11 \notin Output(N')$ means there are no two consecutive 1s in the list $Output(N')$. This theorem is also proved by induction on the structure of the input list.

\subsubsection{The Inhibitor Effect}
The next property is an important one because it has the potential to help us detect inactive zones of the brain. Normally, a human neuron does not have negative weights for all of its inputs but when one or more positive weight inputs are out of order because of some kind of disability, this property can occur. It is called the inhibitor effect because it is important for proving properties of archetype \ref{archetypes}(e). We consider here the single neuron case.  When a neuron has only one input and the weight of that input is less than 0, then the neuron is inactive, which means that for any input, the neuron cannot emit 1 as output. i.e., if a signal reaches this neuron, it will not pass through. As with the other properties, the input sequence has an arbitrary finite length.  This property is expressed as follows.

\begin{theorem}
\label{Prop3} \textbf{[Inhibitor effect]}
$$\forall (N:neuron)  (input:list \; nat),$$
$$ length(w(N))=1 \wedge w_1(N)< 0 \rightarrow 1 \notin Output(N')$$
\end{theorem}

In the statement above, $1 \notin Output(N')$ means there is no 1 in the list $Output(N').$  This property is also proved by induction on the structure of the input.
The inhibitor effect expressed in Theorem \ref{Prop3} has a more general version, which we plan to prove as a future work. For a neuron with multiple inputs, when all input weights are less than or equal to 0, then the neuron is inactive and can not pass any signal. Thus, in addition to proving this property for arbitrary input length, we intend to generalize it to an arbitrary number of neurons.

\subsubsection{The Delayer Effect in a Simple Series}
The next property is about the archetype shown in Figure \ref{archetypes}(a). In this structure, each neuron output is the input of the next neuron. If we have a series of $n$ single input neurons and all of them have the delayer effect, then the output of the whole structure is the input plus $n$ leading zeros. In other words, this structure transfers the input sequence exactly with a delay marked by the $n$ leading zeros, denoted as $zeros(n)$ in the statement of the theorem below. This theorem is expressed as follows.

\begin{theorem}
\label{Prop4} \textbf{[Delayer effect in a simple series]}
$$\forall (Series: list \; neuron)  (input:list \; nat) (i:nat),$$
$$ length(Series)=n \wedge 0 \leq i < n \wedge{} $$
$$length(w(Series[i]))=1 \wedge w_1 (Series[i])>\tau(Series[i]) $$
$$\rightarrow Output = zeros(n)+ input $$
\end{theorem}

This time, the proof proceeds by induction on the length of $Series$.

\subsection{\label{sec:Neuro-Conclusion}Conclusions and Future Work on the Archetypes Section}

In this section, we have proposed a formal approach to modeling and
validating
leaky integrate and fire neurons and some basic circuits. In the literature, this is not the first attempt to formally investigate neural networks. In \cite{DMGRG16HSB, DLGRG17CSBIO}, the synchronous paradigm is exploited to model neurons and some small neuronal circuits with relevant topological structure and behavior, and to prove some properties concerning their dynamics. Our approach uses the Coq proof assistant
and has turned out to be much more general. As a matter of fact, we guarantee that the properties we prove are true in the general case, such as true for any input values, any length of input, and any amount of time. As an example, let us consider the simple series. In \cite{DLGRG17CSBIO}, the authors were able to write a function (more precisely, a Lustre node) which encodes the expected behavior of the circuit. Then, they could call a model checker to test whether the property at issue is valid for some input series with a fixed length. Here we can prove that the desired behavior is true for any length of series and any parameters.

We plan several future works. As a first next step, we intend to formally study the missing
archetypes of Figure \ref{archetypes} (series with multiple outputs,
parallel composition, negative loop, and contralateral inhibition) and
other new archetypes made of two, three or more neurons. We have already started to investigate the two-neuron positive loop, where the first neuron activates the second one, which in turn activates the first one. Our progress so far includes defining a Coq record that expresses the structure of the positive loop, along with an inductive predicate that relates the two neurons and their corresponding two lists of values obtained by applying the potential function over time. This predicate is true whenever the output has a particular pattern that is important for proving one of the more advanced properties we are studying.  Defining general relations that can be specialized to specific patterns will likely also be very useful for the kinds of properties that are important for more complex networks.

As a second next step, we plan to focus on the composition of the studied archetypes. There are two main ways to couple two circuits: either to connect the output of the first one to the input of the second one, or to nest the first one inside the second one. We are interested in detecting the compositions which lead to circuits with a meaningful biological behavior. Archetypes can be considered as the syllables of a given alphabet. When two or more syllables are combined, it is possible to obtain either a real word or a word which does not exist. In the same way, the archetype composition may or may not lead to meaningful networks.  As a long-term aim, with the help of the neurophysiologist Franck Grammont, we would like to be able to prove that whatever neural network can be expressed as a combination of the small mini-circuits we have identified, similar to how all the words can be expressed as a combination of the syllables of a given alphabet. We believe that the power of our theorem proving approach will allow us to advance rapidly in the study of the fundamental structural and functional properties of the elementary building blocks of the brain and cognition.




\section{Conclusion and Perspective} \label{sec:conclusion}

{\bf In the field of biomedicine and biology in general.} 
The ongoing revolution in AI is accelerating the development of
software that enables computers to perform ``intelligent'' clinical and medical tasks. 
Machine learning algorithms find hidden patterns in data, classify and associate similar
patients/diseases/drugs based on common features (e.g., the IBM Watson system which is used to analyse genomic and cancer data).  
Future challenges in medicine include understanding bias in data collection (and also in doctor's experience) and fostering the ability to integrate evidence from heterogeneous datasets, from different omics and clinical data, from several lines of independent data.
We believe that machine learning could satisfy well these needs and
that there is also a need to develop methods that offer a
hypothesis-driven approach, so that doctors do not feel that they are going to be
replaced.  Such methods could provide them with a personalised and easily interpretable
clinical support decision-making tool that could perform a synthesis
of qualitative and quantitative multi-modal evidence. Examples of
decision trees used in current practice for breast cancer diagnosis
can be found at pages 598--603 of \cite{MushlinGreene}.
Our logical approach, although focused on driver mutations, goes in
such a direction and could be used with continuous and discrete mixed
variables.
This information could be obtainable through single cell experiments on cancer biopsies (although with large variance), which is now 
at the stage of passing from basic science to clinical protocols. 
Machine learning could analyse cancer mutation patterns and feed our
logic approach with this information that could be integrated with
other rules such as changes on the metabolic networks or on epigenetics. 
Other rules could be derived from other levels of cancer clinical investigation such as from image data (changes in fMRI, CT-scans and 
microscopy samples), blood analyses (identification and counts of circulating cancer cells) and other types of medical observations. 
The long term plan is to build a portable resource that facilitates diagnostic and therapeutic decision making and promotes a cost-effective personalised patient workup.
This would represent a new paradigm in personalised and precision cancer treatment which integrates multi-modality analyses and 
clinical characteristics in a near-real time manner, improving clinical management of cancer.
Finally we believe that logical approaches could improve the 
harmonisation and standardisation of the reporting and interpretation of clinically relevant data.

The logic approach could have far reaching applications; for example in hematopoiesis each cell branching brings a large number of questions, 
particularly about regulative circuits, experimental settings, departure from homeostasis during diseases or alterations.
Disease conditions could be monitored by generating a score according to a CHESS-(``Changes in Health, End-stage disease, and Signs and Symptoms'') scale. 
CHESS is a summary measure based on a count of comorbidity progression as well as symptoms and clinician ratings of a prognosis of less than six months or so. 
Other applications can be more ambitious. We highlight three of them:

\begin{itemize} 
\item {Explore the dependencies of all human cells types during embryogenesis.} 
This approach could make use of cell atlas \cite{Regev2017} and tissue atlas, see for example \cite{Gamazon2018}.

\item {Explore the dependencies of models of cell dependencies.}
This would lead helping the curation of models databases, see for example \cite{Glont2017},
therefore logic could be used for automatic check of consistency and help in model annotation.

\item {Deriving ontologies.}
The biological information has complex structure and 
it requires an organization that includes controlled vocabularies and formats for the exchange of structured data.
Bio-ontologies are consensus-based, controlled
vocabularies for biological terms and interaction between humans and
computers  \cite{Ashburner2000,Bodenreider2006,Musen2012}. 
Addressing the cell status will require ontologies that take into account the profound wiring of biological processes in the body. 
Here we envision to use logic to automatically generate ontologies from a set of rules.
\end{itemize} 

In summary we believe that together with the current machine learning,
logic could find a 
central position in modeling biomedicine.

{\bf In the neuroscience area.}
Although, the proofs we have completed require some sophisticated reasoning, there is still a significant amount that is common between them.
As we continue, we expect to encounter more complex inductions as we consider more complex properties.
Thus, it will become important to automate as much of the proofs as possible, most likely by writing tactics tailored to the kind of induction, case analysis, and mathematical reasoning that is needed here.
Furthermore, defining general relations that can be specialized to
specific patterns will likely also be very useful for the kinds of
properties that are important for more complex networks.

So far, logic approaches turned out to be particularly suited to investigate
dynamic properties of some canonical neuronal circuits,
and we believe they will be crucial to
formally study the behavior of bigger circuits obtained by archetype composition. 


{\bf More generally, in biology.} 
We believe that logic will allow both the integration of scalability (i.e. from neurons to brain superior abilities) 
and a better use of unstructured data, 
such as users and doctors insights in the form of experience and personal judgement. 
Moreover explanability and interpretability will allow a team-in-the loop approach to medical cases, 
i.e. analysis of patients that could involve at different levels nurses, single doctors or teams of specialists and consultants. 
This interaction will allow a logic clinical decision system to re-interpret findings on the lights of additional medical expertise and event outcomes handling. 
This aspect will be a key stone in Intensive Care Unit (ICU) when physicians are asked to act quickly and in an explanable modus.

We believe that, because of the above mentioned properties, logic could be used in clinical trials and medical protocols. 
An important component that would be needed to help team (human-in the loop) scalability in addressing diseases is to
generate a visualisation of real-time hypothesis and logic solutions 
allowing to interpret the results according to 
the personalisable features selected for.



\bibliographystyle{splncs04}
\bibliography{references,neuro} 


%

\end{document}